\documentclass[10pt,fleqn]{article}

\usepackage{latexsym}
\usepackage{amsbsy}
\usepackage{amsfonts}
\usepackage{amssymb}
\usepackage{amscd}
\usepackage{t1enc}
\usepackage{epsfig}
\usepackage{color}
\usepackage{cite}

\mathchardef\Gamma="0100
\mathchardef\Delta="0101
\mathchardef\Theta="0102
\mathchardef\Lambda="0103
\mathchardef\Xi="0104
\mathchardef\Pi="0105
\mathchardef\Sigma="0106
\mathchardef\Upsilon="0107
\mathchardef\Phi="0108
\mathchardef\Psi="0109
\mathchardef\Omega="010A

\def\I{{\rm i}}
\def\D{{\rm d}}
\def\E{{\rm e}}

\def\vec{\boldsymbol}

\def\cal{\mathcal}

\input{diagxy}

\begin{document}

\title{ Effect of Diffraction on Wigner Distributions  of 
    Optical  Fields and how to Use It in Optical Resonator Theory.\\ II -- Unstable Resonators}

 \date{}

  \maketitle
\begin{center}


  \vskip -1cm

{
\renewcommand{\thefootnote}{}
{\bf   Pierre Pellat-Finet\footnote{pierre.pellat-finet@univ-ubs.fr,
  eric.fogret@univ-ubs.fr} and \'Eric
  Fogret}
}
\setcounter{footnote}{0}

\medskip
{\sl \small Laboratoire de Mathématiques de Bretagne Atlantique UMR CNRS 6205

Université de Bretagne Sud, B. P. 92116, 56321 Lorient cedex, France}
\end{center}

\vskip 1cm

\begin{center}

\begin{minipage}{12cm}
\hrulefill

\smallskip
{\small
{\bf Abstract.} The second part of the article is devoted to field transfers by
diffraction that are represented by fractional Fourier transformations
whose orders are complex numbers. The corresponding effects on the
Wigner distributions associated with optical fields are still represented by
 $4\times 4$ matrices operating on the scaled phase-space, but unlike matrices involved in the first part, those  matrices  decompose into two matrices that
essentially represent  2--dimensional hyperbolic rotations, not elliptical rotations. The result is applied to the theory of
 unstable resonators.

\smallskip
\noindent {\sl Keywords:} Diffraction, Fourier optics,   fractional-order Fourier
transformation, unstable optical resonators, spherical angular spectrum,  Wigner distribution.

\smallskip

\noindent{\sl PACS:} 42.30.Kq

\smallskip
\noindent {\bf Content}

\smallskip

\noindent 1. Introduction \dotfill \pageref{sect1}

\noindent 2. Field transfer by
  diffraction:  complex-order transfer \dotfill \pageref{sect2}

\noindent 3. Complex scaled angular-variables \dotfill \pageref{sect3}

\noindent 4. Effect of diffraction on Wigner distributions: complex-order transfers \dotfill \pageref{sect4}

\noindent 5. Application to unstable optical resonators\dotfill \pageref{sect5}

\noindent 6. Conclusion\dotfill \pageref{conc}

\noindent Appendix A\dotfill \pageref{appenA2}

\noindent Appendix B\dotfill \pageref{appenB}

\noindent Appendix C\dotfill \pageref{appenC}


\noindent  References\dotfill \pageref{refe2}

}

\hrulefill
\end{minipage}
\end{center}

\section{Introduction}\label{sect1}

Diffraction phenomena considered in the first part of the paper are
represented by fractional Fourier transformations whose orders are real
numbers (see Part I \cite{Part1}). In the scaled phase-space, the effect of diffraction on the Wigner distribution of an optical
field  is then expressed by a $4\times 4$ matrix which
splits into two matrices representing pure (or elliptical) rotations operating on  two 2-dimensional disjoint subspaces. 
If the field transfer between two mirrors of an
(open) optical
resonator corresponds to such a diffraction phenomenon---associated with a real-order fractional Fourier transformation--- the
resonator is said to be stable, and usual properties of such a resonator can be deduced from the
invariance of  Wigner distributions associated with  the resonator
transverse modes \cite{Part1}.

Sometimes, a real fractional-order cannot, indeed,  be associated with
a given diffraction phenomenon, and in completing the theory, we 
introduce complex orders \cite{PPF5,Fog1,Fog3}; that is done in this part. Using
complex orders leads us to define complex scaled variables; nevertheless, the
method we  employ remains similar to the one developed in the first part. Pure
rotations of the first part are changed into hyperbolic rotations. 
More precisely, the effect of a given diffraction phenomenon on Wigner
distributions is represented by a $4\times 4$ matrix which splits into
two matrices corresponding to 
hyperbolic rotations, plus an elliptical rotation in some cases,
operating on appropriate 2--dimensional disjoint subspaces of the scaled
phase-space, as will be shown.

If the field transfer between two mirrors of an optical resonator is
expressed by a complex-order fractional Fourier transformation, the
resonator is said to be unstable \cite{PPF5,Fog1,Fog3}, as confirmed by analyzing how Wigner distributions
behave in the transfer.

\section{Field transfer by diffraction:  complex-order transfer}\label{sect2}

\subsection{Complex order associated with a diffraction phenomenon}\label{sect21}


Once more we consider  the field transfer from a spherical emitter ${\cal
  A}_1$ (curvature radius $R_1$) to a spherical
receiver ${\cal A}_2$ (radius $R_2$) at a distance $D$. The field
  amplitude $U_2$ on ${\cal A}_2$ is related to the field amplitude
  $U_1$ on ${\cal A}_1$ by  
  Eq. (I.2)\footnote{Equation ($n$) of part I is referred as Eq.\ (I.n).} of the first part, that is,
\begin{eqnarray}
U_2(\vec r')\!\!\!\!&=&\!\!\!\!{\I\over \lambda D} \exp\left[-{\I\pi\over
 \lambda}\left({1\over R_2}+{1\over D}\right)\vec r'\vec \cdot\vec
 r'\right] \label{eq1} \\
& & \hskip 2cm \times \;\int_{{\mathbb R}^2}\exp\left[-{\I\pi\over
 \lambda}\left({1\over D}-{1\over R_1}\right)\vec r\vec\cdot\vec r\right]\, 
\exp\left({2\I\pi\over \lambda D}\,\vec r\vec\cdot\vec
r'\right)\,U_1(\vec r)\, \D\vec r\,,\nonumber 
\end{eqnarray}
where $r^2$ and $r'^2$ have been replaced respectively by the Euclidean scalar
products $\vec
r\vec\cdot\vec r$ and $\vec r'\!\vec\cdot\vec r'\!$, which will be
more convenient for generalizing to complex scaled-variables.

Let $f$ be a function of a two-dimensional real variable. We recall that the 2--dimensional fractional Fourier transform of order $\alpha$ of  $f$ is defined by
\begin{equation}
  {\cal F}_\alpha[f](\vec \rho ')={\I\E^{\I\alpha}\over \sin\alpha}\exp (-\I\pi \vec\rho'\vec\cdot\vec\rho '\cot\alpha )\int_{{\mathbb R}^2}\!\!\!\exp (-\I\pi \vec \rho \vec\cdot\vec\rho \cot\alpha )\,\exp\left(2\I\pi \vec\rho '\vec\cdot\vec\rho\over\sin\alpha\right)\,f(\vec\rho )\,\D\vec\rho\,,\label{eq2n}\end{equation}
where $\alpha$ may be a complex number \cite{Nam,Mcb}.

To express the right-hand member of Eq.\ (\ref{eq1}) by using a fractional-order Fourier transformation, 
we look at the parameter $J$ such that
\begin{equation}
J={(R_1-D)(R_2+D)\over D(D-R_1+R_2)}\,.\label{eq3}\end{equation}

The case $J\ge 0$ is considered in the first part of the
paper and corresponds to real-order transfers, since the associated
parameter $\alpha$ is a real number.

In this part, we assume $J<0$, and since
complex $\alpha$ will be used, we say that the transfer from ${\cal
  A}_1$ to ${\cal A}_2$ is a ``complex-order'' transfer.
We first note that $J=-1$ is not
realistic (it corresponds to $R_1=0$ or $R_2=0$) and will not be considered. Then, when $J<0$, we choose $\alpha$ as follows.
\begin{enumerate}
\item If $J<-1$, let $\beta$ be the real  number whose sign is the sign of $D$ (then $\beta D>0$), and
  such that $\coth^2\beta =-J$. We choose $\alpha =
  \I \beta$, so that $\cot\alpha =-\I \coth \beta$.
\item If $-1<J<0$,  let $\beta$ be the real number  whose sign is the sign of $D$, and such that  $\coth^2\beta
  =-1/J$.
  To obtain $\cot\alpha
  =-\I/\coth\beta$,  we choose $\alpha$  as follows:
\begin{itemize}
\item If $D>0$, then $\alpha =\pi/2+\I\beta$. (Remark: $\beta >0$.)
\item If $D<0$, then $\alpha =-\pi /2+\I\beta$. (Remark: $\beta <0$.)
\end{itemize}
\end{enumerate}
In every case (items 1 and 2 above), we have $\cot^2\alpha =J$, a result similar to the one given by Eq.$\!$~(I.5) in part~I, but now $\alpha$ is a complex number.

\subsection{Complex scaled-variables and complex scalar-product}\label{sect22}
To express Eq.\ (\ref{eq1}) by using a fractional-order Fourier transformation, that is, in the form of Eq.~(\ref{eq2n}), and to take an example, we have to match $\exp (-\I \pi \vec \rho\vec\cdot\vec\rho\cot\alpha)$ with $\exp (-\I\pi A\,\vec r\vec\cdot\vec r)$, where $A$ is a real number. If the fractional order is chosen to be $\alpha =\I\beta$ ($\beta$ a real number, see above),  we notice that $\I\cot\alpha =\coth\beta$ is a real number, and we have to reintroduce a complex quantity in $\exp (-\I \pi \vec \rho\vec\cdot\vec\rho\cot\alpha)$. That is done by using complex scaled-variables, namely, replacing two-dimensional real vectors $\vec\rho$ and $\vec \rho '$ by two-dimensional complex vectors, as will be shown. The same can be done when $\alpha =\pm(\pi /2)+\I\beta$.

On the other hand, the dot product used in Eqs.\ (\ref{eq1}) and (\ref{eq2n})  is  a two--dimensional Euclidean scalar product, defined  for real vectors $\vec \rho = (\rho_x,\rho_y)$ and $\vec \rho '=(\rho_x',\rho_y')$  (where $\rho_x$, $\rho_y$, $\rho_x'$ and $\rho_y'$ are real numbers) by
\begin{equation}
  \vec \rho\vec \cdot \vec \rho '=\rho_x\rho_x'+\rho_y\rho_y'\,.\end{equation}
A  complex vector, say $\vec \sigma$, is written  $\vec\sigma =(\sigma_x,\sigma_y)=(r_x+\I s_x,r_y+\I s_y)$ where $r_x$, $r_y$, $s_x$ and $s_y$ are real numbers. The complex scalar product of vectors $\vec\sigma =(\sigma_x,\sigma_y)=(r_x+\I s_x,r_y+\I s_y)$ and 
 $\vec\sigma ' =(\sigma'_x,\sigma'_y)=(r'_x+\I s'_x,r'_y+\I s'_y)$ is defined by
\begin{equation}
  \vec \sigma \!\vec\cdot\!\vec\sigma '=\sigma_x\sigma'_x+\sigma_y\sigma'_y
  =r_xr'_x+r_yr'_y-s_xs'_x-s_ys'_y+\I (r_xs'_x+r_ys'_y+r'_xs_x+r'_ys_y)\,.
\end{equation}
Indeed, it is a symmetric bilinear form (it is not an Hermitian scalar product) and  a generalization of the previous Euclidean scalar product.

\subsection{Comparing signs of $R_1(R_1-D)$ and $R_2(R_2+D)$}\label{sect23}

The signs of $R_1(R_1-D)$ and $R_2(R_2+D)$ will be useful both for the definition of appropriate scaled variables and fractional parameters, and for expressing some results.
From the identity
$D(D-R_1+R_2)=R_1R_2-(R_1-D)(R_2+D)$, 
we deduce
\begin{equation}
{1\over J}={D(D-R_1+R_2)\over (R_1-D)(R_2+D)}={R_1R_2\over
  (R_1-D)(R_2+D)}-1\,,\end{equation} 
that is,
\begin{equation}
{R_1R_2\over
  (R_1-D)(R_2+D)}=1+{1\over J}\,.\end{equation}
We conclude as follows:
\begin{enumerate}
\item If $J<-1$, then $-1<1/J<0$, so that
\begin{equation}
{R_1R_2\over
  (R_1-D)(R_2+D)}>0\,,\end{equation}
and $R_1(R_1-D)$ and $R_2(R_2+D)$ have the same sign.
\item If $-1<J<0$, then $1/J<-1$, and 
\begin{equation}
{R_1R_2\over
  (R_1-D)(R_2+D)}<0\,,\end{equation}
which means that  $R_1(R_1-D)$ and $R_2(R_2+D)$ have opposite signs. %
\end{enumerate}

\subsection{Field transfer for $D>0$ and $J<-1$}

\subsubsection{Complex scaled-variables and scaled field-amplitudes ($D>0$ and $J<-1$)}\label{sect231}


Let $\frak s$ be the sign of $R_1(R_1-D)$ and let
\begin{equation}
  \chi_1 ={\frak s}{D\over R_1-D}\coth\beta\,.\label{eq10}\end{equation}
Since $\beta D > 0$, we have $D\coth\beta >0$, and then
\begin{equation}
\chi_1 R_1={\frak s}{R_1D\over R_1-D}\coth\beta >0\,.\label{eq11}\end{equation}
Let $\chi_2$ be such that
\begin{equation}
\chi_2 ={\frak s}{D\over R_2+D}\coth\beta\,.\label{eq12}\end{equation}
Since $R_1(R_1-D)$ and $R_2(R_2+D)$ have the same sign, and since  $D\coth\beta >0$, we obtain
\begin{equation}
\chi_2R_2= {\frak s} {R_2D\over R_2+D}\coth\beta
>0\,.\label{eq12b}\end{equation}
Introducing ${\frak s}$ in Eq.\ (\ref{eq10}) is a way for obtaining $\chi_1R_1>0$ (and then $\chi_2R_2>0$), a condition that will be useful later.

Finally, we note that $\varepsilon_1$ and $\varepsilon_2$ defined by
\begin{equation}
\varepsilon_1 =-{\frak s}\,\I\,\chi_1 =-\I {D\over R_1-D}\coth\beta ={D\over
  R_1-D}\cot\alpha\,,\end{equation}
and
\begin{equation}
\varepsilon_2 =-{\frak s}\,\I \,\chi_2 =-\I {D\over R_2+D}\coth\beta ={D\over
  R_2+D}\cot\alpha\,,\end{equation}
are complex extensions of $\varepsilon_1$ and $\varepsilon_2$, defined in 
  the first part of the paper: with respect to $\alpha$, they are as in the real case---see Eqs.\ (I.7) and (I.8).

To define complex scaled-variables, we proceed in two steps. Since $\chi_1R_1>0$ and $\chi_2R_2>0$, we first
introduce real scaled-variables $\vec \rho$ on ${\cal A}_1$ and $\vec
\rho '$ on ${\cal A}_2$, such that
\begin{equation}
\vec \rho ={\vec r\over \sqrt{\lambda\chi_1R_1}}\,,\hskip .5cm
\mbox{and}\hskip .5cm
\vec \rho ' ={\vec r'\over
  \sqrt{\lambda\chi_2R_2}}\,,\label{eq13}
\end{equation}
and scaled field-amplitudes
\begin{equation}
V_1(\vec \rho )=\sqrt{\chi_1R_1\over \lambda}\,U_1\left(\sqrt{\lambda\chi_1R_1}\,\vec \rho
\right)\,,\label{eq18}
\end{equation}
and 
\begin{equation}
V_2(\vec \rho ')=\sqrt{\chi_2R_2\over \lambda}\,U_2\left(\sqrt{\lambda\chi_2R_2}\,\vec \rho '
\right)\,.\label{eq18b}
\end{equation}
Then we define complex scaled-variables $\vec\sigma$ on ${\cal A}_1$ and $\vec \sigma
'$ on ${\cal A}_2$ by
\begin{equation}
\vec \sigma ={1+{\frak s}\I \over \sqrt{2}}\,\vec\rho ={1+{\frak s}\I \over
  \sqrt{2\lambda\chi_1R_1}}\,\vec r\,,\label{eq14}\end{equation}
and
 \begin{equation}
\vec \sigma ' ={{\frak s}+\I\over \sqrt{2}} \,\vec\rho' ={{\frak s}+\I \over
  \sqrt{2\lambda\chi_2R_2}}\,\vec r'\,.\label{eq14b}\end{equation}
In the following, we  denote $\Gamma =(1+{\frak s}\I ){\mathbb R}$ and $\Gamma '=({\frak s}+\I ){\mathbb R}$, so that $\vec \sigma \in\Gamma\times \Gamma$ and $\vec \sigma '\in\Gamma '\times\Gamma '$.
  
The corresponding scaled amplitudes, defined on $\Gamma\times \Gamma$ and $\Gamma '\times\Gamma '$ respectively, are
\begin{equation}
V_{{\rm c}1}(\vec \sigma )=V_1\left( {1-{\frak s}\I\over \sqrt{2}}\,\vec \sigma \right)=\sqrt{\chi_1R_1\over \lambda}\,U_1\left({1-{\frak
    s}\I\over \sqrt{2}}\sqrt{\lambda\chi_1R_1}\,\vec\sigma\right)\,,\label{eq15}\end{equation}
and
\begin{equation}
V_{{\rm c}2}(\vec \sigma ')=V_2\left({{\frak s}-\I\over \sqrt{2}}\,\vec\sigma '\right) =\sqrt{\chi_2R_2\over \lambda}\,U_2\left({{\frak
    s}-\I\over \sqrt{2}}\sqrt{\lambda\chi_2R_2}\,\vec\sigma
    '\right)\,.\end{equation}
(Index c indicates that $V_{{\rm c}1}$  and $V_{{\rm c}2}$ are defined
for complex variables.)

\subsubsection{Explicit expression of the field transfer by a hyperbolic fractional-order Fourier transform ($D>0$, $J<-1$)}
   By using the previous scaled variables, scaled functions and complex scalar product, we prove in Appendix A that for $\vec \sigma '\in \Gamma '\times\Gamma '$, Eq.\ (\ref{eq1}) can be written
\begin{equation}
V_{{\rm c}2}(\vec \sigma ')= {\I\,{\frak s}\over\sin\alpha}\exp
(-\I\pi\vec\sigma '\!\!\vec\cdot\vec\sigma '\cot\alpha )
 \!\!\int_{\Gamma\times \Gamma} \!\!\!\!\!\!\!\exp (-\I\pi\vec\sigma \vec\cdot\vec
\sigma\cot\alpha )
\exp\!\left({2\I\pi\over\sin\alpha}\vec\sigma\! \vec\cdot\!\vec\sigma
  '\!\right)V_{{\rm c}1}(\vec\sigma )\,\D\vec \sigma \,,\label{eq20n}\end{equation}
where $\alpha$ is chosen as in Sect. \ref{sect21}: $\alpha =\I\beta$ ($\beta$ a real number with $\beta D>0$).

Apart from a constant factor, the right-hand part of Eq.\ (\ref{eq20n}) is formally identical to the fractional Fourier transform of order $\alpha$ of the amplitude $V_{{\rm c}1}$---see Eq.~(\ref{eq2n})---but variables are (two-dimensional) complex variables and the integration domain is $\Gamma\times\Gamma$ in place of ${\mathbb R}^2$. The image domain (to which $\vec\sigma '$ belongs) is $\Gamma '\times \Gamma '$. It is in this sense, and with some abuse, that the field transfer is said to be expressed by a fractional-order Fourier transform; and since the order is a complex number, the field transfer is called a complex-order transfer.

Nevertheless, in practice, we prefer to use real scaled variables, which will be helpful in introducing Wigner distributions on a scaled phase-space identical to the one used in Part I.
Indeed, integration in  Eq.\ (\ref{eq20n}) can be achieved on ${\mathbb R}^2$ in place of $\Gamma\times \Gamma$ changing $\vec \sigma$ into $\vec\rho =(1-{\frak s}\I)\vec\sigma /\sqrt{2}$. For $\vec\rho '=({\frak s}-\I)\vec\sigma ' /\sqrt{2}$, 
Eq.\ (\ref{eq20n}) becomes
\begin{eqnarray}
    V_{2}(\vec \rho ')&=& {{\I}\over\sinh\beta}\exp
    (-\I\,{\frak s}\,\pi\vec\rho '\vec\cdot\vec\rho '\coth\beta ) \nonumber \\
    & & \hskip .3cm\times\;\int_{{\mathbb R}^2} \!\!\exp (-\I\,{\frak s}\,\pi\vec\rho\vec\cdot\vec
\rho\coth\beta )
\exp\!\left({2\I\,\pi\over\sinh\beta}\vec\rho \vec\cdot\vec\rho
'\right)V_{1}(\vec\rho )\,\D\vec \rho\,.\label{eq20p}\end{eqnarray}
Since $\vec\rho$ and $\vec\rho '$ are real vectors, we  have $\vec\rho\vec\cdot\vec\rho=||\vec \rho ||^2=\rho^2$ and $\vec\rho '\vec\cdot\vec\rho '=||\vec \rho '||^2=\rho'^2$.

We define the ``hyperbolic fractional Fourier transform'' of order $\beta$ ($\beta \in{\mathbb R}$) of function $f$ by
\begin{equation}
    \cal {H}_\beta[f](\vec \rho ')= {\I\E^{-\beta}\over\sinh\beta}\exp
(-\I\pi{\rho '}^2\coth\beta )
 \!\!\int_{{\mathbb R}^2} \!\!\!\!\exp (-\I\pi{\rho}^2\coth\beta )
\exp\!\left({2\I\pi\over\sinh\beta}\vec\rho \vec\cdot\vec\rho
  '\!\right)\,f(\vec\rho )\,\D\vec \rho\,,\label{eq20q}\end{equation}
so that, for ${\frak s}=1$,  Eq.\ (\ref{eq20p}) becomes
\begin{equation}
    V_{2}(\vec \rho ')=\E^{\beta}\,\cal{H}_{\beta} [V_1](\vec \rho ')\,,\label{eq17a}\end{equation}
and for ${\frak s}=-1$
\begin{equation}
  V_{2}(\vec \rho ')=-\E^{-\beta}\,\cal{H}_{-\beta} [V_1](-\vec \rho ')\,,\label{eq17b}\end{equation}

Equations (\ref{eq17a}) and (\ref{eq17b}) are synthetized in
\begin{equation}
  V_{2}(\vec \rho ')={\frak s}\,\E^{{\frak s}\beta}\,\cal{H}_{{\frak s}\beta} [V_1]({\frak s}\vec \rho ')\,,\label{eq17}\end{equation}
which is similar to Eq.\ (I.12) apart that  it expresses the field transfer from ${\cal A}_1$ to ${\cal A}_2$ by a first kind of ``hyperbolic fractional-order Fourier transformation''.

\subsection{Field transfer for $D<0$ and $J<-1$}

For $D<0$, the field transfer from ${\cal A}_1$ to ${\cal A}_2$ is virtual. We define $\alpha=\I\beta$ with $\beta <0$ and $\coth^2\beta = -J$, and we show  in Appendix B
\begin{equation}
V_{{\rm c}2}(\vec \sigma ')= {{\frak s}\,\I\over\sin\alpha}\exp
(\I\pi\vec\sigma '\!\!\vec\cdot\vec\sigma '\cot\alpha )
 \!\!\int_{\Gamma\times \Gamma} \!\!\!\!\!\!\!\exp (\I\pi\vec\sigma \vec\cdot\vec
\sigma\cot\alpha )
\exp\!\left(-{2\I\pi\over\sin\alpha}\vec\sigma\! \vec\cdot\!\vec\sigma
  '\!\right)V_{{\rm c}1}(\vec\sigma )\,\D\vec \sigma \,,\label{eq20s}\end{equation}
for appropriate complex scaled-variables (given in Appendix B). Formally, up to a multiplicative factor,  we have a fractional Fourier transformation of order $-\alpha$.

By changing complex scaled variable into real one, we obtain
\begin{eqnarray}
    V_{2}(\vec \rho ')&=& {{\I}\over\sinh\beta}\exp
    (\I\,{\frak s}\,\pi\vec\rho '\vec\cdot\vec\rho '\coth\beta ) \nonumber \\
    & & \hskip .3cm\times\;\int_{{\mathbb R}^2} \!\!\exp (\I\,{\frak s}\,\pi\vec\rho\vec\cdot\vec
\rho\coth\beta )
\exp\!\left({2\I\,\pi\over\sinh\beta}\vec\rho \vec\cdot\vec\rho
'\right)V_{1}(\vec\rho )\,\D\vec \rho\,,\label{eq20t}\end{eqnarray}
which can be written as
\begin{equation}
  V_{2}(\vec \rho ')=-{\frak s}\,\E^{-{\frak s}\beta}\,\cal{H}_{-{\frak s}\beta} [V_1](-{\frak s}\vec \rho ')\,,\label{eq17c}\end{equation}
which is similar to Eq.\ (\ref{eq17}).

\subsection{Field transfer for $D>0$ and $-1<J<0$}

\subsubsection{An additional condition}
In the previous sections, since $\alpha =\I\beta$, we had $\I/\sin\alpha =1/\sinh\beta$, so that using complex scaled-variables according to Eqs. (\ref{eq14}) and (\ref{eq14b}) reintroduced a factor $\I$ in $2\I\pi \vec\sigma\vec\cdot\vec\sigma '/\sin\alpha$; this factor was necessary to match with the factor $2\I\pi \vec r\vec \cdot\vec r' /\lambda D$ of Eq.\ (\ref{eq1}).

For $-1<J<0$ and $D>0$, the fractional order is chosen to be $\alpha =\pi /2+\I\beta$, with $\beta>0$.
Since $\cot\alpha =-\I/\coth\beta$, matching for example $\I\pi \vec \rho\vec\cdot\vec \rho \cot\alpha$ with $\I\pi A\,\vec r\vec\cdot\vec r$ (where $A$ is a real number) leads us to introduce complex vectors. But here, we have
$\sin\alpha =\cos \I\beta =\cosh\beta$, which is a real number, and the previous scaled variables do not allow matching $2\I\pi \vec\sigma\vec\cdot\vec\sigma '/\sin\alpha$ with $2\I\pi \vec r\vec \cdot\vec r' /\lambda D$. This is why we will use different complex scaled-variables.

\subsubsection{Complex scaled-variables and scaled field-amplitudes ($D>0$, $-1<J<0$)}
Let ${\frak s}$ still denote the sign of  $R_1(R_1-D)$. 
We define
\begin{equation}
\chi_1={{\frak s}D\over R_1-D}\,{1\over \coth\beta}\,,\label{eq20}\end{equation}
and since $D$ and $\beta$ have the same sign, we have
\begin{equation}
  \chi_1R_1={\frak s}{R_1D\over R_1-D}\,{1\over \coth\beta} >0\,.\label{eq21}\end{equation}
We then define $\chi_2$ by
\begin{equation}
\chi_2=-{\frak s}{D\over R_2+D}\,{1\over \coth\beta}\,,\label{eq22}\end{equation}
and since the sign of
$R_2(R_2+D)$ is opposite to the sign of $R_1(R_1-D)$, we obtain
\begin{equation}
  \chi_2R_2=-{\frak s}{R_2D\over R_2+D}\,{1\over \coth\beta} >0\,.\label{eq23}\end{equation}
Finally, we define 
\begin{equation}
\varepsilon_1 =-{\frak s}\,\I\,\chi_1 =-\I {D\over R_1-D}{1\over \coth\beta} ={D\over
  R_1-D}\cot\alpha\,,\end{equation}
and
\begin{equation}
\varepsilon_2 ={\frak s}\,\I\,\chi_2 =-\I {D\over R_2+D}{1\over \coth\beta} ={D\over  R_2+D}\cot\alpha\,,\end{equation}
which are, with respect to $\alpha$, as in the real case.

We use $\vec \rho$ and $\vec \rho'$ as in
Eq.\ (\ref{eq13}) and scaled field amplitudes $V_1$ and $V_2$ as in Eqs.\
(\ref{eq18}) and (\ref{eq18b}). Then we define complex scaled variables on ${\cal A}_1$
and ${\cal A}_2$ by
\begin{equation}
  \vec \sigma ={1+\I\over \sqrt{2}}\,\vec\rho ={1+\I \over \sqrt{2\lambda\chi_1R_1}}\,\vec r\,,\label{eq38a}\end{equation}
and \begin{equation}
\vec \sigma ' ={1-\I \over \sqrt{2}}\,\vec \rho '= {1-\I \over
  \sqrt{2\lambda\chi_2R_2}}\,\vec r'\,.\label{eq39a}\end{equation}
The corresponding scaled amplitudes are
\begin{equation}
V_{{\rm c}1}(\vec \sigma )=\sqrt{\chi_1R_1\over \lambda}\;U_1\!\left({1-\I\over \sqrt{2}}\sqrt{\lambda\chi_1R_1}\;\vec\sigma\right)\,,\end{equation}
and
\begin{equation}
V_{{\rm c}2}(\vec \sigma ')=\sqrt{\chi_2R_2\over \lambda}\;U_2\!\left({1+\I\over \sqrt{2}}\sqrt{\lambda\chi_2R_2}\;\vec\sigma
    '\right)\,.\end{equation}

\subsubsection{Field-amplitude transfer}
By using the previous scaled vectors and scaled field amplitudes,  we obtain that Eq.\ (\ref{eq1}) can be written as
\begin{eqnarray}
V_{{\rm c}2}(\vec \sigma ')&=& {\I \over\sin\alpha}\exp
(-{\frak s}\,\I\pi\vec\sigma '\vec\cdot \vec\sigma '\cot\alpha )\nonumber \\
& &\hskip 1cm \times
\int_{{\Gamma ''}\times {\Gamma ''}} \!\!\!\!\!\exp (-{\frak s}\,\I\pi\vec\sigma\!\vec\cdot\!\vec
\sigma\cot\alpha )
\exp\!\left({2\I\pi\over\sin\alpha}\vec\sigma\! \vec\cdot\!\vec\sigma
'\!\right)V_{{\rm c}1}(\vec\sigma )\,\D\vec \sigma \,,\label{eq20nb}\end{eqnarray}
where $\Gamma ''=(1+\I ){\mathbb R}$ and where $\vec \sigma '\in (1-\I){\mathbb R}\times (1-\I){\mathbb R}$.
The proof is given in Appendix C.

Formally, Eq.\ (\ref{eq20nb}) involves (up to a multiplicative factor)  a fractional Fourier transformation defined on $\Gamma ''\times \Gamma ''$, whose  order is $\alpha$ or $\pi -\alpha$.

For studying the effect of diffraction on Wigner distributions, we use real variables, according to Eqs.\ (\ref{eq38a}) and (\ref{eq39a}),  and write Eq.\ (\ref{eq20nb}) in the form
\begin{equation}
  V_2(\vec \rho ')=-
  {1\over
  \cosh\beta}\,\exp\, \left({{\frak s}\,\I\pi \rho '^2\over \coth\beta}\right) 
\int_{{\mathbb R^2}} \!\!\!\exp\, \left(-{{\frak s}\,\I \pi \rho^2\over \coth\beta}\right) 
\,\exp\left({2\I\pi \vec \rho\vec \cdot\vec\rho '\over
  \cosh\beta}\right)\,V_1(\vec \rho )\,\D\vec\rho\,.\label{eq39}
\end{equation}

We define a second kind of ``hyperbolic fractional Fourier transformation'' of order $\beta$ ($\beta \in{\mathbb R}$) by
\begin{equation}
    \cal {K}_\beta[f](\vec \rho ')= {\I\E^{-\beta}\over\cosh\beta}\exp
\left({\I\pi{\rho '}^2\over \coth\beta} \right)
 \!\!\int_{{\mathbb R}^2} \!\!\!\!\exp \left(- {\I\pi{\rho}^2\over \coth\beta} \right)
\exp\!\left({2\I\pi \vec\rho \vec\cdot\vec\rho
  '\over\cosh\beta}\right)\,f(\vec\rho )\,\D\vec \rho\,,\label{eq20qb}\end{equation}
so that 
Eq.\ (\ref{eq39}) can be written
\begin{equation}
  V_2(\vec \rho ')= \I\E^{{\frak s}\beta}{\cal K}_{{\frak s}\beta}[V_1](\vec \rho ')\,.\label{eq45}
  \end{equation}

\section{Complex scaled angular-variables}\label{sect3}

If $U$ denotes the field amplitude on a spherical cap, the corresponding spherical angular-spectrum is \cite{Part1}
\begin{equation}
  S(\vec \Phi)={1\over \lambda^2}\,\widehat{U}\left({\vec \Phi\over\lambda}\right)\,,\end{equation}
where $\vec \Phi$ denotes the angular spatial-frequency, related to the spatial frequency $\vec F$ by
$\vec\Phi =\lambda \vec F$.

The transfer of the spherical spectrum by diffraction is governed by the same laws as that of the field amplitude  \cite{PPF4b}. This holds true for the scaled angular-spectrum, namely, Eqs.\ (\ref{eq17}), (\ref{eq17c}) and (\ref{eq45}) hold true if scaled field-amplitudes $V_1$ and $V_2$ are replaced by the scaled spherical angular spectra $\widehat V_1$ and $\widehat V_2$, and scaled spatial-variables are replaced by scaled angular-frequencies (see Eqs. (I.12) and (I.24) \cite{Part1}).

We now provide the scaled angular-variables corresponding to the  scaled spatial-variables of Sect.\ \ref{sect2}. They are helpful in expressing the transfer of the spherical angular-spectrum and we will use some of them in the third part of the article.  
In defining scaled angular-variables,
 we manage to preserve Eqs.\ (I.21--22). First,
we introduce real scaled angular-variables on ${\cal A}_1$ and ${\cal
  A}_2$ according to
\begin{equation}
\vec\phi =\sqrt{\chi_1R_1\over \lambda}\,\vec \Phi \,,\hskip .5cm \mbox{
  and}\hskip .5cm
\vec\phi ' =\sqrt{\chi_2R_2\over \lambda}\,\vec \Phi '\,. 
\end{equation}
Then complex scaled angular-variables on ${\cal A}_1$ (denoted $\vec\theta$)
and ${\cal A}_2$  ($\vec\theta '$) are defined as follows.
\begin{itemize}
\item If $J<-1$, we choose
  \begin{equation}
\vec\theta ={1-{\frak s}\I\over \sqrt{2}}\,\vec \phi
={\sqrt{\chi_1R_1\over 2\lambda }}(1-{\frak s}\I)\,\vec \Phi\,,\end{equation}
  and
  \begin{equation}
\vec\theta '= {{\frak s}-\I\over \sqrt{2}}\,\vec \phi '=
{\sqrt{\chi_2R_2\over 2\lambda}} ({\frak s}-\I )\,\vec \Phi '\,.\label{eq30}
\end{equation}
\item If $-1<J<0$, we choose
  \begin{equation}
\vec\theta ={1-\I\over \sqrt{2}}\,\vec \phi
={\sqrt{\chi_1R_1\over 2\lambda }}(1-\I)\,\vec \Phi\,,\end{equation}
  and
  \begin{equation}
\vec\theta '= {1+\I\over \sqrt{2}}\,\vec \phi '=
\sqrt{\chi_2R_2\over 2\lambda} (1+\I )\,\vec \Phi '\,.
\end{equation}
(Scaled variables $\vec\phi$, $\vec\phi'$, $\vec\theta$ and $\vec\theta '$ are 2-dimensional (vectorial) variables.)
\end{itemize}

In both cases we obtain
\begin{equation}
\vec r\vec \cdot \vec F={1\over \lambda} \vec r\vec \cdot \vec \Phi
=\vec\rho\vec\cdot\vec\phi =\vec\sigma\vec\cdot\vec\theta\,,
\end{equation}
and 
\begin{equation}
\vec r'\vec \cdot \vec F'={1\over \lambda} \vec r'\vec \cdot \vec \Phi'
=\vec\rho'\vec\cdot\vec\phi '=\vec\sigma'\vec\cdot\vec\theta',
\end{equation}
which generalize Eqs.\ (I.21--22) to complex scaled-variables.

\section{Effect of diffraction on Wigner distributions:  complex-order transfers}\label{sect4}

\subsection{Hyperbolic rotations}

In ${\mathbb R}^2$, the hyperbolic rotation  of parameter
$\beta$ (a real number) transforms the point $P=(x,y)$ into the point $P'=(x',y')$
such that
\begin{equation}
\pmatrix{x'\cr y'\cr}=
\pmatrix{\cosh \beta &\sinh\beta \cr \sinh\beta &\cosh\beta\cr}
\pmatrix{x\cr y\cr}\,.\label{eq31}
\end{equation}
We also call ``angle'' of the hyperbolic rotation the parameter $\beta$.

\begin{figure}[h]
\begin{center}
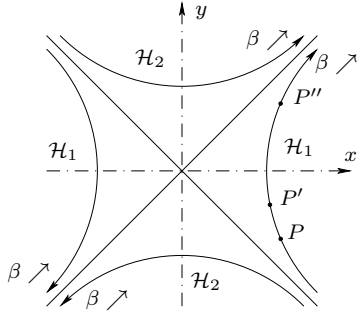
\caption{\small Equilateral hypberbolas ${\cal H}_1$ (equation $x^2-y^2=A>0$)
  and ${\cal H}_2$ (equation $y^2-x^2=A>0$). Each hyperbola has two
  branches. Let $P$ be a point on ${\cal H}_1$. By applying
  successive hyperbolic rotations, $P$ becomes $P'$, then $P''$, etc. and all these points remain on the same branch of the same hyperbola. Arrows indicate how hyperbola branches
  are run for increasing values of the rotation angle (denoted $\beta$).\label{fig0}}
\end{center}
\end{figure}

Consider a hyperbolic rotation of angle $\beta$ and a point $P=(x,y)$. Let $A={x}^2-{y}^2$ and assume $A\ne 0$. Eq.\ (\ref{eq31}) leads to
${x'}^2-{y'}^2= {x}^2-{y}^2=A$, which means that point $P$ and
its image $P'$ in the previous hyperbolic rotation
belong to the equilateral hyperbola ${\cal H}$, whose equation is $x^2-y^2= A$, and whose
 asymptotes are the bisectors of the $x$ and
 $y$--axes. For $A>0$, the hyperbola ${\cal H}$ is as ${\cal H}_1$ in Fig.\ \ref{fig0}; and for $A<0$, it is as ${\cal H}_2$.

Since 
\begin{equation}
\pmatrix{\cosh \beta ' &\sinh\beta ' \cr \sinh\beta ' &\cosh\beta '\cr}
\pmatrix{\cosh \beta &\sinh\beta \cr \sinh\beta &\cosh\beta\cr}=
\pmatrix{\cosh (\beta +\beta ') &\sinh (\beta +\beta ') \cr \sinh
  (\beta +\beta ') &\cosh (\beta +\beta ')\cr}
\,,\label{eq32}
\end{equation}
the (commutative) composition of two hyperbolic rotations, with respective angles
$\beta$ and $\beta '$,  is the hyperbolic rotation of angle $\beta
+\beta '$.

Let $P=(x,y)\ne (0,0)$ and $P'=(x',y')$ be as in Eq.\ (\ref{eq31}), and
let $P''=(x'',y'')$ be the image of $P'$ in the hyberbolic
rotation of angle $\beta '$. Then points $P$, $P'$ and $P''$
are on the previous hyperbola ${\cal H}$, and $P''$ is the image of
$P$ by the hyperbolic rotation whose angle is $\beta +\beta '$ (see Fig.\ \ref{fig0}, where both $\beta$ and $\beta '$ are positive).

By applying successive hyperbolic rotations, we obtain a sequence of
points that belong to the same branch of a same hyperbola.

\subsection{Diffraction and Wigner distribution}

\subsubsection{Wigner distribution}
We use the scaled phase-space  related to real scaled-variables. The Wigner distribution associated with a scaled field-amplitude $V$ is defined by
\begin{equation}
  W(\vec \rho ,\vec \phi )=\int_{{\mathbb R}^2}V\left(\vec \rho +{\vec\tau\over 2}\right)
  \,\overline{V\left(\vec \rho -{\vec\tau\over 2}\right)}\,\exp (2\I\pi \vec\phi\vec\cdot\vec \tau )\,\D\tau\,.\end{equation}

In the following, we consider a spherical  emitter ${\cal A}_1$ and a spherical receiver ${\cal A}_2$. The Wigner distribution associated with the field amplitude on ${\cal A}_j$ ($j=1,2)$ is denoted $W_j$.

We will show that the result obtained in Part I for real $\alpha$ (Eq.\ (I.30), Sect.\ 5) can be extended to complex $\alpha$, with $\alpha =\I\beta$ or $\alpha =\pm (\pi /2)+\I\beta$.

\subsubsection{Transfer of the Wigner distribution for $J<-1$ and ${\frak s}=1$}

\subsubsection*{\sl Transformation expression}
We have $\alpha =\I\beta$. 
We will show that the equivalent of Eq.\ (I.30) takes the form
\begin{equation}
  W_{2}(\vec\rho,\vec\phi)=W_{1}(\vec\rho \cosh\beta -\vec\phi\sinh\beta  ,-\vec\rho\sinh\beta +\vec\phi\cosh\beta )\,,\label{eq50}\end{equation}
which means that the elliptical rotation involved in Eq.\ (I.30)  is replaced by a hyperbolic one, as we will explain.

\smallskip
\noindent {\sl Proof.}
We define $E(x)=\exp (\I \pi x)$,  as in Part I. We consider Eq.\ (\ref{eq20p}) with ${\frak s}=1$  and obtain
\begin{eqnarray}
W_2(\vec \rho, \vec \phi )\!\!\!\! &=&
    \int_{{\mathbb R}^2}V_{2}\left(\vec\rho +{\vec\tau\over 2}\right)\,\overline{V_2\left(\vec\rho  -{\vec\tau\over 2}\right) }\,\exp (2\I\pi \vec\phi \vec\cdot\vec\tau )\,\D\vec\tau
    \nonumber \\
    &=&\!\!\!\!{1\over \sinh^2\!\beta}
\int_{{\mathbb R}^2} E \left(-\left\Vert \vec \rho +{\vec \tau  \over 2}\right\Vert^2\coth\beta\right)\nonumber\\
&&\times  \left\{\int_{{\mathbb R}^2} \!\! E(-\rho
'^2\coth\beta)\,E\!\left[{2\vec\rho
    '\over\sinh\beta}\!\vec\cdot\!\left(\vec\rho+{\vec\tau\over
    2}\right)\right]\, V_1(\vec\rho ')\,\D\vec\rho '\right. \nonumber \\
&&\times \;E\left(\left\Vert \vec \rho -{\vec
  \tau\over 2}\right\Vert^2\!\!\coth\beta\right)
\nonumber \\
&&\times\left.
 \int_{{\mathbb R}^2}E(\rho ''^2\coth\beta)\; E\left[-{2\vec\rho ''\over\sinh\beta}\vec\cdot\left(\vec\rho-{\vec\tau\over 2}\right)\right] \,\overline{V_1(\vec \rho '')}\,\D \vec \rho ''\right\} \,E(2\vec\tau\vec\cdot\vec\phi  )\,\D\vec \tau\, \nonumber \\
&=&\!\!\!\!{1\over \sinh^2\!\beta}
\int_{{\mathbb R}^2} E(-\rho'^2\coth\beta )\,E\!\left({2\vec \rho\vec\cdot\vec\rho '\over \sinh\beta}\right)\,V_1(\vec\rho ')\,\D\vec\rho '\nonumber \\
&&\times \;\int_{{\mathbb R}^2}\!\!\!E(\rho''^2\coth\beta )\,E\!\left(-{2\vec \rho\vec\cdot\vec\rho ''\over \sinh\beta}\right)\,\overline{V_1(\vec\rho '')}\,\D\vec\rho ''\nonumber \\
&&\times \;\int_{{\mathbb R}^2}\!\!E(-2\vec\rho\vec\cdot\vec\tau\coth\beta )
\,E\!\left({\vec\rho '+\vec\rho ''\over\sinh\beta}\vec\cdot\vec\tau\right)
\,E(2\vec\tau\vec\cdot\vec\phi )\,\D\vec\tau\,.
\label{eq32t}
\end{eqnarray}
If $\delta$ denotes the Dirac generalized function, the last integral in Eq. (\ref{eq32t}) is equal to
\begin{equation}
\delta \left(\vec\phi -\vec\rho\coth\beta +{\vec\rho '+\vec\rho ''\over 2\sinh\beta}\right)
 =4\sinh^2\!\beta\;\delta \bigl(2\vec\phi\sinh\beta -2\vec\rho\cosh\beta +\vec\rho '+\vec\rho ''\bigr)\,,
\end{equation}
so that Eq. (\ref{eq32t}) becomes
\begin{eqnarray}
W_2(\vec \rho, \vec \phi )\!\!\!\!&=&\!\!\!\!
4\int_{{\mathbb R}^2}\!\!E(-\rho'^2\coth\beta )\,E\!\left({2\vec\rho\vec\cdot\vec\rho '\over\sinh\beta}\right) E\bigl(\Vert -2\vec\rho\cosh\beta -2\vec\phi\sinh\beta -\vec\rho '\Vert^2\coth\beta \bigr)\nonumber 
 \\
&& \times \;E\left[-{2\vec \rho\over\sinh\beta}\vec\cdot \bigl(2\vec\rho\cosh\beta -2\vec\phi\sinh\beta -\vec\rho '\bigr)\right]
V_1(\vec\rho ')\nonumber \\
& & \times\; \overline{V_1(2\vec\rho\cosh\beta -2\vec \phi\sinh\beta -\vec\rho ')}\,\D\vec\rho '\,.
 \label{eq33n}
\end{eqnarray}
We change $\vec\rho '$ into
$\vec \tau =2\vec \rho '-2\vec\rho \cosh\beta+2\vec\phi\sinh\beta$, so that\footnote{Remember that both $\vec\tau$ and $\vec\rho '$ are real 2--dimensional variables. If $\vec \tau =(\tau_x,\tau_y)=2(\rho_x',\rho'_y)=2\vec\rho '$, then $\D\vec \tau =\D\tau_x\,\D\tau_y=4\,\D\rho'_x\,\D\rho'_y=4\,\D\vec \rho '$.} $\D\vec\tau =4\,\D\vec \rho '$, and
 Eq. (\ref{eq33n}) becomes
\begin{eqnarray}
W_2(\vec \rho, \vec \phi )\!\!\!\! &=&\!\!\!\!\int_{{\mathbb R}^2}
V_1\left(\vec\rho \cosh\beta -\vec\phi\sinh\beta +{\vec\tau\over 2}\right)
 \overline{V_1\left(\vec\rho \cosh\beta -\vec\phi\sinh\beta
   -{\vec\tau\over 2}\right)}\nonumber  \\
& & \hskip 6cm \times\;
 E\bigl[2(\vec\phi \cosh\beta -\vec\rho\sinh\beta )\vec\cdot\vec\tau\bigr] \,\D\vec \tau
 \nonumber \\
 &=& W_1(\vec\rho \cosh\beta -\vec\phi\sinh\beta, -\vec\rho\sinh\beta +\vec\phi \cosh\beta)\,.
\end{eqnarray}
which is Eq. (\ref{eq50}). The proof is complete.

\subsubsection*{\sl Matrix expression} 
\begin{figure}[h]
  \begin{center}
    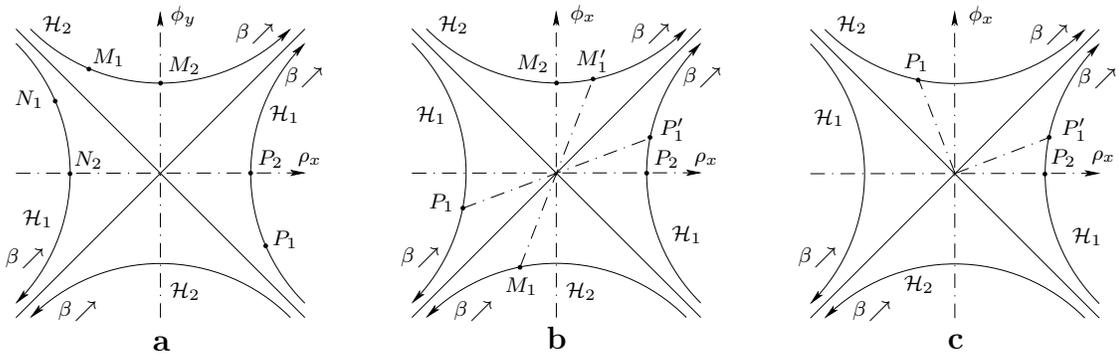
 \caption{Effect of diffraction in the scaled subspace $\rho_x$--$\phi_x$ (the same would hold in the $\rho_y$--$\phi_y$ subspace). The value taken by $W_2$ at  $P_2$ is equal to the value taken by $W_1$ at $P_1$. (a) $\alpha =\I\beta$, $\beta >0$, ${\frak s}=1$; the effect of diffraction is a hyperbolic rotation of angle $\beta$. (b)  $\alpha =\I\beta$, $\beta >0$, ${\frak s}=-1$; the effect of diffraction is a $\pi$--rotation followed by a hyperbolic rotation of angle $-\beta$. (c) $\alpha =(\pi /2)+\I\beta$, $\beta >0$, ${\frak s}=1$; the effect of diffraction is a $-\pi/2$--rotation followed by a hyperbolic rotation of angle $-\beta$. \label{fig2.4}}
  \end{center}
\end{figure}

According to Equation (\ref{eq50}), the value taken by  $W_2$ at point $(\vec\rho ,\vec\phi)$ is the value taken by $W_1$ at point $(\vec\rho \cosh\beta -\vec\phi\sinh\beta  ,-\vec\rho\sinh\beta +\vec\phi\cosh\beta )$. In the subspace $\rho_x$--$\phi_x$, this corresponds to a hyperbolic rotation of angle $\beta$. To understand that, consider the value taken by $W_2$ at point $P_2=(\rho_x,\phi_x)=(1,0)$, which is equal to the value taken by  $W_1$  at point $P_1=(\cosh\beta , -\sinh\beta)$. For $\beta >0$, the point $P_2$ is deduced from $P_1$ as shown in Fig.\ \ref{fig2.4}--a, that is, in the hyperbolic rotation of angle $\beta$.

The same conclusion is obtained by considering point $M_2=(0,1)$, which is deduced from point  $M_1=(-\sinh\beta ,\cosh\beta)$, and point $N_2=(-1,0)$, which comes from $N_1=(-\cosh\beta ,\sinh\beta)$.

In the $\rho_x$--$\phi_x$ plane, the matrix expression of the corresponding  hyperbolic rotation is  
\begin{equation}
\pmatrix{\rho'_x\cr\phi_x'}=\pmatrix{\cosh\beta & \sinh\beta\cr
\sinh\beta &\cosh\beta}\pmatrix{\rho_x\cr
  \phi_x}\,,\label{eq37a}\end{equation}
whose angle is $\beta$.

The same result is obtained in the $\rho_y$--$\phi_y$, so that the effect of diffraction in the whole scaled-space is a 4--dimensional Wigner rotation 
which can be seen as the product of two hyperbolic rotations  in two 2-dimensional
subspaces. Then Eq.\ (\ref{eq50}) can be written as  a coordinate transformation, whose  matrix form is
\begin{equation}
  \begin{pmatrix}{
    \rho'_x \cr
    \phi'_x\cr
     \rho'_y\cr
    \phi'_y}\end{pmatrix}
  =
  \begin{pmatrix}{\cosh\beta & \sinh\beta & 0 & 0\cr
    \sinh\beta  & \cosh\beta & 0 & 0 \cr
     0 & 0 &\cosh\beta&  \sinh\beta  \cr
    0& 0 & \sinh\beta  &\cosh\beta}\end{pmatrix}
  \begin{pmatrix}{
    \rho_x \cr
    \phi_x\cr
     \rho_y\cr
    \phi_y}\end{pmatrix}\,,\label{eq58}
\end{equation}
that is,
\begin{equation}
  \begin{pmatrix}{
    \rho'_x \cr
    \rho'_y\cr
    \phi'_x\cr
    \phi'_y}\end{pmatrix}
  =
  \begin{pmatrix}{\cosh\beta & 0 &\sinh\beta & 0\cr
    0 & \cosh\beta & 0 & \sinh\beta \cr
    \sinh\beta & 0 & \cosh\beta & 0 \cr
    0& \sinh\beta & 0 &\cosh\beta}\end{pmatrix}
  \begin{pmatrix}{
    \rho_x \cr
    \rho_y\cr
    \phi_x\cr
    \phi_y}\end{pmatrix}
      \,.\label{eq59}
\end{equation}

With $\vec p=(\vec\rho ,\vec\phi )$, Eq.\ (\ref{eq59}) is the matrix form of   $\vec p'={\cal W}\vec p$, where ${\cal W}$  denotes a 4--dimensional Wigner rotation, and  eventually
Eq.\ (\ref{eq50}) takes the form
\begin{equation}
  W_2(\vec p')=W_1({\cal W}^{-1}\vec p')\,,\hskip .5cm \mbox{or}\hskip.5cm
  W_2({\cal W}\vec p)=W_1(\vec p)\,.
\end{equation}

\subsubsection{Transfer of the Wigner distribution for $J<-1$ and ${\frak s}=-1$}

\subsubsection*{\sl Transformation expression}
We  now show that, for $J<-1$ and ${\frak s}=-1$, the transfer by diffraction from ${\cal A}_1$ to ${\cal A}_2$, operates  on the corresponding  Wigner distributions  according to
\begin{equation}
  W_{2}(\vec\rho,\vec\phi)=W_{1}(-\vec\rho \cosh\beta -\vec\phi\sinh\beta  ,-\vec\rho\sinh\beta -\vec\phi\cosh\beta )\,.\label{eq50b}\end{equation}


\bigskip
\noindent {\sl Proof.} By the definition of the Wigner distribution and by Eq.\ (\ref{eq20p}), we obtain,  for ${\frak s}=-1$,
\begin{eqnarray}
W_2(\vec \rho, \vec \phi )\!\!\!\! &=&
    \int_{{\mathbb R}^2}V_{2}\left(\vec\rho +{\vec\tau\over 2}\right)\,\overline{V_2\left(\vec\rho  -{\vec\tau\over 2}\right) }\,\exp (2\I\pi \vec\phi \vec\cdot\vec\tau )\,\D\vec\tau
    \nonumber \\
    &=&\!\!\!\!{1\over \sinh^2\!\beta}
\int_{{\mathbb R}^2} E \left(\left\Vert \vec \rho +{\vec \tau  \over 2}\right\Vert^2\coth\beta\right)\nonumber\\
&&\times  \left\{\int_{{\mathbb R}^2} \!\! E(\rho
'^2\coth\beta)\,E\!\left[{2\vec\rho
    '\over\sinh\beta}\!\vec\cdot\!\left(\vec\rho+{\vec\tau\over
    2}\right)\right]\, V_1(\vec\rho ')\,\D\vec\rho '\right. \nonumber \\
&&\times \;E\left(-\left\Vert \vec \rho -{\vec
  \tau\over 2}\right\Vert^2\!\!\coth\beta\right)
\nonumber \\
&&\times\left.
 \int_{{\mathbb R}^2} E(-\rho ''^2\coth\beta)\; E\left[-{2\vec\rho ''\over\sinh\beta}\vec\cdot\left(\vec\rho-{\vec\tau\over 2}\right)\right] \,\overline{V_1(\vec \rho '')}\,\D \vec \rho ''\right\} \,E(2\vec\tau\vec\cdot\vec\phi  )\,\D\vec \tau\, \nonumber \\
&=&\!\!\!\!{1\over \sinh^2\!\beta}
\int_{{\mathbb R}^2} E(\rho'^2\coth\beta )\,E\!\left({2\vec \rho\vec\cdot\vec\rho '\over \sinh\beta}\right)\,V_1(\vec\rho ')\,\D\vec\rho '\nonumber \\
&&\times \;\int_{{\mathbb R}^2}\!\!\!E(-\rho''^2\coth\beta )\,E\!\left(-{2\vec \rho\vec\cdot\vec\rho ''\over \sinh\beta}\right)\,\overline{V_1(\vec\rho '')}\,\D\vec\rho ''\nonumber \\
&&\times \;\int_{{\mathbb R}^2}\!\!E(2\vec\rho\vec\cdot\vec\tau\coth\beta )
\,E\!\left({\vec\rho '+\vec\rho ''\over\sinh\beta}\vec\cdot\vec\tau\right)
\,E(2\vec\tau\vec\cdot\vec\phi )\,\D\vec\tau\,.
\label{eq32q}
\end{eqnarray}
The last integral in Eq. (\ref{eq32q}) is equal to
\begin{equation}
\delta \left(\vec\phi +\vec\rho\coth\beta +{\vec\rho '+\vec\rho ''\over 2\sinh\beta}\right)
 =4\sinh^2\!\beta\;\delta \bigl(2\vec\phi\sinh\beta +2\vec\rho\cosh\beta +\vec\rho '+\vec\rho ''\bigr)\,,
\end{equation}
so that Eq. (\ref{eq32q}) becomes
\begin{eqnarray}
W_2(\vec \rho, \vec \phi )\!\!\!\!&=&\!\!\!\!
4\int_{{\mathbb R}^2}\!\!E(\rho'^2\coth\beta )\,E\!\left({2\vec\rho\vec\cdot\vec\rho '\over\sinh\beta}\right) E\bigl(-\Vert 2\vec\rho\cosh\beta +2\vec\phi\sinh\beta +\vec\rho '\Vert^2\coth\beta \bigr)\nonumber 
 \\
&& \times \;E\left[{2\vec \rho\over\sinh\beta}\vec\cdot \bigl(2\vec\rho\cosh\beta +2\vec\phi\sinh\beta +\vec\rho '\bigr)\right]
V_1(\vec\rho ')\nonumber \\
& & \times\; \overline{V_1(-2\vec\rho\cosh\beta -2\vec \phi\sinh\beta -\vec\rho ')}\,\D\vec\rho '\,.
 \label{eq33nt}
\end{eqnarray}
We change $\vec\rho '$ into
$\vec \tau =2\vec \rho '+2\vec\rho \cosh\beta+2\vec\phi\sinh\beta$, so that 
 Eq. (\ref{eq33nt}) becomes
\begin{eqnarray}
W_2(\vec \rho, \vec \phi )\!\!\!\! &=&\!\!\!\!\int_{{\mathbb R}^2}
V_1\left(-\vec\rho \cosh\beta -\vec\phi\sinh\beta +{\vec\tau\over 2}\right)
 \overline{V_1\left(-\vec\rho \cosh\beta -\vec\phi\sinh\beta
   -{\vec\tau\over 2}\right)}\nonumber  \\
& & \hskip 6cm \times\;
 E\bigl[2(\vec\phi \cosh\beta +\vec\rho\sinh\beta )\vec\cdot\vec\tau\bigr] \,\D\vec \tau
 \nonumber \\
 &=& W_1(-\vec\rho \cosh\beta -\vec\phi\sinh\beta, -\vec\rho\sinh\beta -\vec\phi \cosh\beta)\,.
\end{eqnarray}
which is eq. (\ref{eq50b}). The proof is complete.

\subsubsection*{\sl Matrix expression} From Eq. (\ref{eq50b}) we conclude that in the subspace $\rho_x$--$\phi_x$ the effect of diffraction is the product of an elliptical rotation of angle $\pi$ and a hyperbolic rotation of angle $-\beta$. To understand that, consider the point $P_2=(1,0)$: according to Eq.\ (\ref{eq50b}), the value taken  by the function $W_2$ at $P_2$ is the value taken by $W_1$ at point $P_1=(-\cosh\beta ,-\sinh\beta)$, as shown in  Fig.\ \ref{fig2.4}--b, for $\beta>0$. Geometrically, $P_1$ is transformed into $P'_1$ ($\pi$--rotation) and $P'_1$ into $P_2$ in the hyperbolic rotation of angle $-\beta$. The same result is obtained from $M_2=(0,1)$, which is the image of $M_1=(-\sinh\beta ,-\cosh\beta )$. The point $M_1$ is transformed into $M'_1$ in a $\pi$--rotation, and $M_2$ is deduced from $M'_1$ in the hyperbolic rotation of angle $-\beta$.
The previous $\pi$-rotation and the hyperbolic rotation commute.

In the $\rho_x$--$\phi_x$ subspace, we have
\begin{equation}
\pmatrix{\rho'_x\cr\phi_x'}
=\pmatrix{\cosh\beta & -\sinh\beta\cr
-\sinh\beta &\cosh\beta}\pmatrix{-1 & 0 \cr 0 &-1}\pmatrix{\rho_x\cr
  \phi_x}=
\pmatrix{-\cosh\beta & \sinh\beta\cr
\sinh\beta &-\cosh\beta}\pmatrix{\rho_x\cr
  \phi_x}\,,\label{eq37b}\end{equation}
The same holds in the $\rho_y$--$\phi_y$ subspace.

Finally, Eq. (\ref{eq50b}) can be written in matrix form as
\begin{equation}
  \begin{pmatrix}{
    \rho'_x \cr
    \phi'_x\cr
     \rho'_y\cr
    \phi'_y}\end{pmatrix}
  =
  \begin{pmatrix}{-\cosh\beta & \sinh\beta & 0 & 0\cr
    \sinh\beta  & -\cosh\beta & 0 & 0 \cr
     0 & 0 &-\cosh\beta&  \sinh\beta  \cr
    0& 0 & \sinh\beta  &-\cosh\beta}\end{pmatrix}
  \begin{pmatrix}{
    \rho_x \cr
    \phi_x\cr
     \rho_y\cr
    \phi_y}\end{pmatrix}\,.
\end{equation}

\subsubsection{Transfer of the Wigner distribution for $-1<J<0$}\label{sect424} 

\subsubsection*{\sl Transformation expression}

For $-1<J<0$, we will prove
\begin{equation}
  W_2(\vec\rho,\vec\phi )=
  W_1(-{\frak s}\,\vec\rho \sinh\beta -\vec\phi\cosh\beta, \vec\rho\cosh\beta +{\frak s}\vec\phi \sinh\beta)\,.
  \label{eq50c}
\end{equation}

\medskip
\noindent {\sl Proof. (i)}
We first provide the proof for ${\frak s}=1$. We use Eq.\ (\ref{eq39}) and, by definition,  we obtain
\begin{eqnarray}
W_2(\vec \rho, \vec \phi )\!\!\!\! &=&
    \int_{{\mathbb R}^2}V_{2}\left(\vec\rho +{\vec\tau\over 2}\right)\,\overline{V_2\left(\vec\rho  -{\vec\tau\over 2}\right) }\,\exp (2\I\pi \vec\phi \vec\cdot\vec\tau )\,\D\vec\tau
    \nonumber \\
    &=&\!\!\!\!{1\over \cosh^2\!\beta}
\int_{{\mathbb R}^2} E \left(\left\Vert \vec \rho +{\vec \tau  \over 2}\right\Vert^2{1\over \coth\beta}\right)\nonumber\\
&&\times  \left\{\int_{{\mathbb R}^2} \!\! E\left(-{\rho
'^2\over \coth\beta}\right)\,E\!\left[{2\vec\rho
    '\over\cosh\beta}\!\vec\cdot\!\left(\vec\rho+{\vec\tau\over
    2}\right)\right]\, V_1(\vec\rho ')\,\D\vec\rho '\right. \nonumber \\
&&\times \;E\left(-\left\Vert \vec \rho -{\vec
  \tau\over 2}\right\Vert^2{1\over \coth\beta}\right)
\nonumber \\
&&\times\left.
 \int_{{\mathbb R}^2} E\left({\rho ''^2\over \coth\beta}\right)\; E\left[-{2\vec\rho ''\over\cosh\beta}\vec\cdot\left(\vec\rho-{\vec\tau\over 2}\right)\right] \,\overline{V_1(\vec \rho '')}\,\D \vec \rho ''\right\} \,E(2\vec\tau\vec\cdot\vec\phi  )\,\D\vec \tau\, \nonumber \\
&=&\!\!\!\!{1\over \cosh^2\!\beta}
\int_{{\mathbb R}^2} E\left(-{\rho'^2\over\coth\beta} \right)\,E\!\left({2\vec \rho\vec\cdot\vec\rho '\over \cosh\beta}\right)\,V_1(\vec\rho ')\,\D\vec\rho '\nonumber \\
&&\times \;\int_{{\mathbb R}^2}\!\!\!E\left({\rho''^2\over \coth\beta}\right)\,E\!\left(-{2\vec \rho\vec\cdot\vec\rho ''\over \cosh\beta}\right)\,\overline{V_1(\vec\rho '')}\,\D\vec\rho ''\nonumber \\
&&\times \;\int_{{\mathbb R}^2}\!\!E\left({2\vec\rho\vec\cdot\vec\tau\over\coth\beta}\right)
\,E\!\left({\vec\rho '+\vec\rho ''\over\cosh\beta}\vec\cdot\vec\tau\right)
\,E(2\vec\tau\vec\cdot\vec\phi )\,\D\vec\tau\,.
\label{eq32qa}
\end{eqnarray}
The last integral in Eq. (\ref{eq32qa}) is equal to
\begin{equation}
\delta \left(\vec\phi +{\vec\rho\over\coth\beta} +{\vec\rho '+\vec\rho ''\over 2\cosh\beta}\right)
 =4\cosh^2\!\beta\;\delta \bigl(2\vec\phi\cosh\beta +2\vec\rho\sinh\beta +\vec\rho '+\vec\rho ''\bigr)\,,
\end{equation}
so that Eq. (\ref{eq32qa}) becomes
\begin{eqnarray}
W_2(\vec \rho, \vec \phi )\!\!\!\!&=&\!\!\!\!
4\int_{{\mathbb R}^2}\!\!E\left(-{\rho'^2\over\coth\beta}\right)\,E\!\left({2\vec\rho\vec\cdot\vec\rho '\over\cosh\beta}\right) E\left(\Vert 2\vec\rho\sinh\beta +2\vec\phi\cosh\beta +\vec\rho '\Vert^2{1\over \coth\beta} \right)\nonumber 
 \\
&& \times \;E\left[{2\vec \rho\over\cosh\beta}\vec\cdot \bigl(2\vec\rho\sinh\beta +2\vec\phi\cosh\beta +\vec\rho '\bigr)\right]
V_1(\vec\rho ')\nonumber \\
& & \times\; \overline{V_1(-2\vec\rho\sinh\beta -2\vec \phi\cosh\beta -\vec\rho ')}\,\D\vec\rho '\,.
 \label{eq33nq}
\end{eqnarray}
We change $\vec\rho '$ into
$\vec \tau =2\vec \rho '+2\vec\rho \sinh\beta+2\vec\phi\cosh\beta$, so that 
 Eq. (\ref{eq33nq}) becomes
\begin{eqnarray}
W_2(\vec \rho, \vec \phi )\!\!\!\! &=&\!\!\!\!\int_{{\mathbb R}^2}
V_1\left(-\vec\rho \sinh\beta -\vec\phi\cosh\beta +{\vec\tau\over 2}\right)
 \overline{V_1\left(-\vec\rho \sinh\beta -\vec\phi\cosh\beta
   -{\vec\tau\over 2}\right)}\nonumber  \\
& & \hskip 6cm \times\;
 E\bigl[2(\vec\rho \cosh\beta +\vec\phi\sinh\beta )\vec\cdot\vec\tau\bigr] \,\D\vec \tau
 \nonumber \\
 &=& W_1(-\vec\rho \sinh\beta -\vec\phi\cosh\beta, \vec\rho\cosh\beta +\vec\phi \sinh\beta)\,.
 \label{eq50d}
\end{eqnarray}
which is Eq. (\ref{eq50c}) for ${\frak s}=1$.

\smallskip \noindent {\sl (ii)}
The proof for  ${\frak s}=-1$ is as follows.   We remark that changing ${\frak s}=1$ into ${\frak s}=-1$ in Eq.\ (\ref{eq39}) is equivalent to changing $\beta$ into $-\beta$. Then the previous derivations  lead to
\begin{eqnarray}
W_2(\vec \rho, \vec \phi )\!\!\!\! &=&\!\!\!\!\int_{{\mathbb R}^2}
V_1\left(\vec\rho \sinh\beta -\vec\phi\cosh\beta +{\vec\tau\over 2}\right)
 \overline{V_1\left(\vec\rho \sinh\beta -\vec\phi\cosh\beta
   -{\vec\tau\over 2}\right)}\nonumber  \\
& & \hskip 6cm \times\;
 E\bigl[2(\vec\rho \cosh\beta -\vec\phi\sinh\beta )\vec\cdot\vec\tau\bigr] \,\D\vec \tau
 \nonumber \\
 &=& W_1(\vec\rho \sinh\beta -\vec\phi\cosh\beta, \vec\rho\cosh\beta -\vec\phi \sinh\beta)\,.
 \label{eq50e}
\end{eqnarray}

\smallskip \noindent {\sl (iii)}
Equations (\ref{eq50d}) and (\ref{eq50e}) are synthetized in Eq. (\ref{eq50c}).

\subsubsection*{\sl Matrix expression}
To obtain the matrix expression of  Eq.\ (\ref{eq50c}), we consider the point $P_2=(1,0)$ in the $\rho_x$--$\phi_x$ subspace. According to Eq.\ (\ref{eq50c}) the value taken by $W_2$ at $P_2$ is the value taken by $W_1$ at point $P_1=(-{\frak s}\sinh\beta ,\cosh\beta )$. Then $P_2$ is deduced from $P_1$ in  rotation of angle $-\pi /2$ followed by a hyperbolic rotation of angle $-{\frak s}\,\beta$, as illustrated in Fig. \ref{fig2.4}--c, for $\beta >0$. In matrix form, we obtain
\begin{equation}
\pmatrix{\rho'_x\cr\phi_x'}=\pmatrix{\cosh\beta & -{\frak s}\,\sinh\beta\cr
-{\frak s}\,\sinh\beta &\cosh\beta}\!\pmatrix{0& 1 \cr -1 &0}\!\!\pmatrix{\rho_x\cr
  \phi_x}
=\pmatrix{{\frak s}\,\sinh\beta & \cosh\beta\cr
-\cosh\beta &-{\frak s}\,\sinh\beta}\!\!\pmatrix{\rho_x\cr
  \phi_x}.\label{eq37c}\end{equation}
The matrix product in Eq.\ (\ref{eq37c}) is not commutative.

The matrix form of Eq. (\ref{eq50c}) is then
\begin{equation}
  \begin{pmatrix}{
    \rho'_x \cr
    \phi'_x\cr
     \rho'_y\cr
    \phi'_y}\end{pmatrix}
  =
  \begin{pmatrix}{{\frak s}\,\sinh\beta & \cosh\beta & 0 & 0\cr
    -\cosh\beta  & -{\frak s}\,\sinh\beta & 0 & 0 \cr
     0 & 0 &{\frak s}\,\sinh\beta&  \cosh\beta  \cr
    0& 0 & -\cosh\beta  &-{\frak s}\,\sinh\beta}\end{pmatrix}
  \begin{pmatrix}{
    \rho_x \cr
    \phi_x\cr
     \rho_y\cr
    \phi_y}\end{pmatrix}
\end{equation}

\bigskip\noindent {\sl Remark.} The matrix in Eq.\ (\ref{eq37c}) is such that
\begin{equation}
\pmatrix{{\frak s}\,\sinh\beta & \cosh\beta\cr
-\cosh\beta &-{\frak s}\,\sinh\beta} = \pmatrix{0& 1 \cr -1 &0}\pmatrix{\cosh\beta & {\frak s}\,\sinh\beta\cr
  {\frak s}\,\sinh\beta &\cosh\beta}\,,\end{equation}
and also corresponds to a hyperbolic rotation of angle ${\frak s}\,\beta$ followed by a rotation of angle $-\pi /2$.

\subsection{Complex rotations}

In the previous section the effect of diffraction on the Wigner distribution associated with an optical field is analyzed in the real scaled phase-space (coordinates $\vec\rho$ and $\vec\phi $). The previous hyperbolic rotations can also be expressed with complex coordinates, as done in the present section.

\subsubsection{Analysis for $J<-1$}
We use $\vec\sigma=(1+{\frak s}\I)\vec\rho /\sqrt{2}$, that is,
\begin{equation}
    \begin{pmatrix}{\rho_x \cr\rho_y}\end{pmatrix}={1-{\frak s}\I \over\sqrt{2}}
      \begin{pmatrix}{\sigma_x \cr \sigma_y}\end{pmatrix}\,,\hskip 1cm
    \begin{pmatrix}{\phi_x \cr\phi_y}\end{pmatrix}={1+{\frak s}\I \over\sqrt{2}}
      \begin{pmatrix}{\theta_x \cr\theta_y}\end{pmatrix}\,,\end{equation}
and
\begin{equation}
    \begin{pmatrix}{\sigma'_x \cr\sigma'_y}\end{pmatrix}={{\frak s}+\I \over\sqrt{2}}
      \begin{pmatrix}{\rho'_x \cr \rho'_y}\end{pmatrix}\,,\hskip 1cm
    \begin{pmatrix}{\theta'_x \cr\theta'_y}\end{pmatrix}={{\frak s}-\I \over\sqrt{2}}
      \begin{pmatrix}{\phi'_x \cr\phi'_y}\end{pmatrix}\,.\end{equation}
  We then obtain
  \begin{eqnarray}
    \begin{pmatrix}{
      \sigma '_x \cr
      \theta '_x\cr }\end{pmatrix}&=&{1\over \sqrt{2}}
    \begin{pmatrix} {{\frak s}+\I & 0  \cr
       0&{\frak s}-\I }
    \end{pmatrix}
     \begin{pmatrix}{
      \rho'_x \cr
      \phi'_x\cr
     }\end{pmatrix}
     \nonumber \\
     &=& {1\over \sqrt{2}}
    \begin{pmatrix} {{\frak s}+\I & 0  \cr
       0&{\frak s}-\I }
    \end{pmatrix}
   \begin{pmatrix}{{\frak s}\cosh\beta & \sinh\beta \cr
       \sinh\beta & {\frak s}\cosh\beta         }
      \end{pmatrix}
     \begin{pmatrix}{
      \rho _x \cr
      \phi _x\cr}\end{pmatrix}
     \nonumber \\
      &=& {1\over 2}
    \begin{pmatrix} {{\frak s}+\I & 0  \cr
       0&{\frak s}-\I }
    \end{pmatrix}
   \begin{pmatrix}{{\frak s}\cosh\beta & \sinh\beta \cr
       \sinh\beta & {\frak s}\cosh\beta         }
   \end{pmatrix}
   \begin{pmatrix}{1-{\frak s}\I & 0 \cr
       0& 1+{\frak s}\I
       }\end{pmatrix}
     \begin{pmatrix}{
      \sigma_x \cr
      \theta_x\cr}\end{pmatrix}
     \nonumber \\
     &=& \begin{pmatrix}{\cosh\beta & \I \sinh\beta \cr
       -\I\sinh\beta &\cosh\beta}
   \end{pmatrix} \begin{pmatrix}{
      \sigma_x \cr
      \theta_x\cr}\end{pmatrix}
      \nonumber \\
     &=& \begin{pmatrix}{\cos\alpha & \sin\alpha \cr
       -\sin\alpha &\cos\alpha}
   \end{pmatrix} \begin{pmatrix}{
      \sigma_x \cr
      \theta_x\cr}\end{pmatrix}\,,\label{eq82}
  \end{eqnarray}
  where $\alpha = \I\beta$. The same can be written with $\rho_y$ and $\phi_y$ so that
 \begin{equation}
    \begin{pmatrix}{
      \sigma '_x \cr
      \theta '_x\cr
      \sigma '_y\cr
      \theta '_y}\end{pmatrix}=
    \begin{pmatrix} {\cos\alpha & \sin\alpha & 0 & 0 \cr
       -\sin\alpha& \cos\alpha  & 0& 0 \cr
       0 & 0 & \cos\alpha & \sin\alpha\cr
      0 & 0 & -\sin\alpha & \cos\alpha}
    \end{pmatrix}
     \begin{pmatrix}{
      \sigma _x \cr
      \theta_x\cr
      \sigma'_y\cr
      \theta'_y}\end{pmatrix}
     \,,\label{eq84}\end{equation}
 which is similar to Eq.\ (I.37).

\subsubsection{Analysis for $-1<J<0$}

 We obtain
  \begin{eqnarray}
    \begin{pmatrix}{
      \sigma '_x \cr
      \theta '_x\cr }\end{pmatrix}&=&{1\over \sqrt{2}}
    \begin{pmatrix} {1-\I & 0  \cr
       0&1+\I }
    \end{pmatrix}
     \begin{pmatrix}{
      \rho'_x \cr
      \phi'_x\cr
     }\end{pmatrix}
     \nonumber \\
     &=& {1\over \sqrt{2}}
    \begin{pmatrix} {1-\I & 0  \cr
       0&1+\I }
    \end{pmatrix}
   \begin{pmatrix}{{\frak s}\sinh\beta & \cosh\beta \cr
       -\cosh\beta & -{\frak s}\sinh\beta         }
      \end{pmatrix}
     \begin{pmatrix}{
      \rho _x \cr
      \phi _x\cr}\end{pmatrix}
     \nonumber \\
      &=& {1\over 2}
    \begin{pmatrix} {1-\I & 0  \cr
       0&1+\I }
    \end{pmatrix}
   \begin{pmatrix}{{\frak s}\sinh\beta & \cosh\beta \cr
       -\cosh\beta & -{\frak s}\sinh\beta         }
   \end{pmatrix}
   \begin{pmatrix}{1-\I & 0 \cr
       0& 1+\I
       }\end{pmatrix}
     \begin{pmatrix}{
      \sigma_x \cr
      \theta_x\cr}\end{pmatrix}
     \nonumber \\
     &=& \begin{pmatrix}{-{\frak s}\,\I \sinh\beta & \cosh\beta \cr
       -\cosh\beta &-{\frak s}\,\I \sinh\beta}
   \end{pmatrix} \begin{pmatrix}{
      \sigma_x \cr
      \theta_x\cr}\end{pmatrix}\,.
  \end{eqnarray}
  We introduce $\alpha =(\pi /2)+\,\I\,\beta$ so that $\cos\alpha =\sin\I\beta =\,\I\,\sinh \beta$ and $\sin\alpha =\cosh\beta$,  and we obtain
  \begin{equation}
     \begin{pmatrix}{
      \sigma '_x \cr
      \theta '_x\cr }\end{pmatrix} =
     \begin{pmatrix}{{\frak s}\cos\alpha & \sin\alpha \cr
      -\sin\alpha &{\frak s}\cos\alpha}
   \end{pmatrix} \begin{pmatrix}{
         \sigma_x \cr
         \theta_x} \end{pmatrix}\,.
  \end{equation}
  For ${\frak s}=1$, we have a rotation of angle $-\alpha$, as in Eq.\ (\ref{eq82}). For ${\frak s}=-1$, we write
\begin{equation}
  \begin{pmatrix}{
      \sigma '_x \cr
      \theta '_x\cr }\end{pmatrix} =
     \begin{pmatrix}{-\cos\alpha & \sin\alpha \cr
      -\sin\alpha &-\cos\alpha}
   \end{pmatrix} \begin{pmatrix}{
         \sigma_x \cr
         \theta_x} \end{pmatrix}
     =
     \begin{pmatrix}{\cos\alpha ' & -\sin\alpha '\cr
      \sin\alpha '&\cos\alpha '}
   \end{pmatrix} \begin{pmatrix}{
         \sigma_x \cr
         \theta_x} \end{pmatrix}\,,
  \end{equation}
which is a rotation of angle $\alpha' =\alpha -\pi =- (\pi /2)+\I\beta$.

\section{Application to unstable optical resonators}\label{sect5}

\subsection{Direct and back transfers in a resonator}\label{sect51}

We consider an optical resonator made up of two spherical mirrors
${\cal M}_1$ (object radius  $R_1$ and image radius $R'_1$) and ${\cal M}_2$ (radii $R_2$ and
$R'_2$).  The algebraic measure from $\Omega_1$ (the vertex of ${\cal M}_1$) to $\Omega_2$ (the vertex of ${\cal M}_2$) is $D=\overline{\Omega_1\Omega_2}$, and it is $D'=\overline{\Omega_2\Omega_1}$ from $\Omega_2$ to $\Omega_1$. Since algebraic measures are positive if taken in the sense of light propagation, which changes after a reflection, we have $D=D'$ and we use the algebraic length of the
resonator, which is $L=D=D'$. (For definitions of object and image radii, see Sect.\ 7.1 Part I; for a definition of the algebraic length $L$, see Sect.\ 7.2, Part I \cite{Part1}.)

For the field transfer from ${\cal M}_1$ to ${\cal M}_2$, the emitter
is ${\cal M}_1$ (image radius $R'_1$) and the receiver is ${\cal M}_2$ (object radius $R_2$), so that by Eq.\ (\ref{eq3}) we obtain
\begin{equation}
J={(R'_1-L)(R_2+L)\over L(L-R'_1+R_2)}\,.
\end{equation}
For the field transfer from ${\cal M}_2$ to ${\cal M}_1$, the emitter
is ${\cal M}_2$  (image radius $R'_2$) and the receiver is ${\cal M}_1$ (object radius $R_1$) and we compute
\begin{equation}
J'={(R'_2-L)(R_1+L)\over L(L-R'_2+R_1)}\,,
\end{equation}
Since $R'_1=-R_1$ and $R'_2=-R_2$, we obtain $J=J'$.

We conclude that the direct and back transfers in a resonator are of the same kind: they are both real-order transfers, or both complex-order transfers.  Moreover, if the order of the transfer from ${\cal M}_1$ to ${\cal M}_2$ is $\beta$, and $\beta '$ for the transfer from ${\cal M}_2$ to ${\cal M}_1$, since $L$ has the same sign for both transfers, we have $\beta =\beta '$.

\subsection{Interpretation in the scaled phase-space for $J=J'<-1$ and ${\frak s}=1$}\label{sect52}

If we use complex scaled-variables, for
  $J=J'<-1$, the field transfer from ${\cal M}_1$ to ${\cal M}_2$
is represented
by a fractional Fourier transformation whose order is $\alpha =\I\beta$, and
the field transfer from ${\cal M }_2$ to ${\cal M}_1$
by a fractional Fourier transformation whose order is $\alpha '=\I\beta '$,
with $\coth^2 \beta = J =J '=\coth^2\beta '$. Since both
$\beta$ and $\beta '$ have the sign of $L$, we have $\beta =\beta '$.
If we use real scaled-variables, to which the scaled phase-space is referred, and according to Eqs.\ (\ref{eq17}) and (\ref{eq17c}), the field transfer is expressed by a hyperbolic fractional Fourier transform whose order is $\pm \beta$.

\begin{figure}
\begin{center}
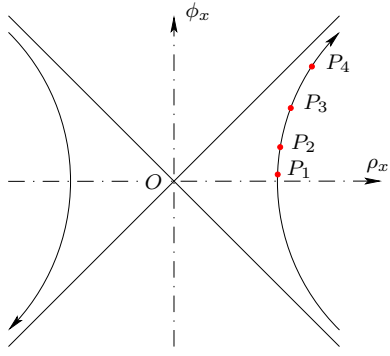
\caption{\small A sequence ($P_i$) for $J=J'<-1$, $\beta >0$ and
  ${\frak s}=1$. The values taken at points $P_1$, $P_3$, $P_5$ etc. by the Wigner distribution on ${\cal M}_1$ are equals. They are also equals to the values of the Wigner distribution on ${\cal M}_2$ taken at points $P_2$, $P_4$, etc.  Point $P_{i+1}$ is deduced from $P_i$ in a hyperbolic rotation of angle $\beta$. If
  $d_i=OP_i$, both sequences $(d_{2i})$ and $(d_{2i+1})$ are  diverging, which shows that  the Wigner distribution on each mirror  spreads over an increasingly wide area after every reflection: the resonator is unstable.\label{fig1}
}
\end{center}
\end{figure}

The interpretation of how Wigner distributions behave in an unstable resonator is carried out in the scaled phase-space and is as follows. We
first consider the  matrix of Eq.\ (\ref{eq58}), that is ${\frak s}=1$. We analyze the situation in the $\rho_x$--$\phi_x$ subspace, in which the effect of diffraction is expressed by Eq.\ (\ref{eq37a}).
We consider a point $P_1=(\rho_{1x},\phi_{1x})\ne (0,0)$ and the hyperbola
whose equation is ${\rho_x}^2-{\phi_x}^2={\rho_{1x}}^2-{\phi_{1x}}^2=A$, to which $P_1$ belongs. The value taken at point $P_1$ by the Wigner distribution on ${\cal M}_1$   equals the value  the Wigner distribution on ${\cal M}_2$ takes at point $P_2$ that is deduced from $P_1$ in the hyperbolic rotation
 of parameter $\beta$. The back transfer from ${\cal M}_2$ to  ${\cal M}_1$  is
 expressed by a hyperbolic rotation of parameter $\beta '=\beta$, which transforms $P_2$ into $P_3$. The value taken at $P_3$ by the Wigner distribution on ${\cal M}_1$ is equal to the value taken at $P_2$ by  the Wigner distribution on ${\cal M}_2$, namely, to the value taken at $P_1$ by the Wigner distribution on ${\cal M}_1$. The point $P_3$ is deduced from $P_1$  in a
 hyperbolic rotation of parameter $2\beta$. If $\beta >0$, the sequence of points $P_i$  corresponds to
 increasing values of $\beta$ (see Fig. \ref{fig1}).  If $d_i=OP_i$, the
 sequence $(d_i)$ is diverging
 (see Fig.\ \ref{fig1}). The same result holds in the $\rho_y$--$\phi_y$ subspace. Eventually, the resonator is unstable, since the support of the Wigner distribution on each mirror  spreads over an increasingly wide area after every reflection. This can also be understood by considering light rays in such a resonator, as will be done in Part III. 

 The same analysis can be  done for $\beta <0$.

 \medskip
 \noindent{\sl Remark.} General properties of unstable resonators are described by Anan'ev \cite{Ana} and  also by Siegman \cite{Sie2}. Qualifying optical resonators as stable or unstable is conventional, but no judicious, since many laser whose cavities  are  unstable resonators perfectly work.
The difference between the two kinds of resonators can be done according to the behaviors of Wigner distributions. In stable resonators, Wigner distributions undergo elliptical rotations so that, after every reflection on a mirror,  the luminous energy remains near the optical axis; these stable resonators are sometimes called ``confined-mode resonators.'' On the contrary, in unstable resonators, Wigner distributions undergo hyperbolic rotations and the energy spread over wider and wider areas after  reflections.

 \subsection{Interpretation in the scaled phase-space for $J=J'<-1$ and ${\frak s}=-1$}

\begin{figure}[h]
\begin{center}
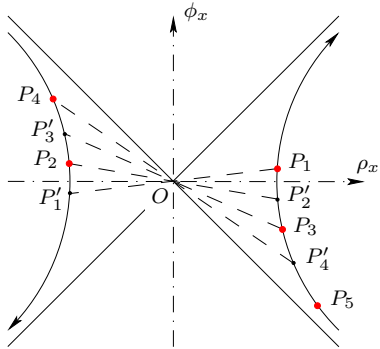
\caption{\small The case $J=J'<-1$, $\beta >0$ and ${\frak s}=-1$. Points $P_i$ are defined  as in Fig.\ \ref{fig1}. Point $P'_i$ is deduced from $P_i$ in a $\pi$--rotation, and $P_{i+1}$ is deduced from $P'_i$ in a hyperbolic rotation of angle $-\beta$.
  If $d_i=OP_i$, the sequences
  $(d_{2i})$ and $(d_{2i+1})$ are diverging. The resonator is unstable. \label{fig2}
}
\end{center}
\end{figure}

In this section, points $P_1$, $P_2$, etc.\ have the same interpretations as in the previous section. We start with point $P_1$.
If ${\frak s}=-1$, according to
Eq.\ (\ref{eq37b}), $P_2$
that can be deduced from $P_1$ as follows: point $P_1$ (and more
generally point $P_i$) undergoes an
elliptic rotation of angle $\pi$ and becomes $P'_1$ ($P'_i$) on the other
branch of the hyperbola (see Fig.\ \ref{fig2}). Then $P'_1$ ($P'_i$) undergoes the hyperbolic rotation of
parameter (angle) $-\beta$ and becomes $P_2$ ($P_{i+1})$ (see Fig.\ \ref{fig2}, drawn for positive $\beta$). We obtain a sequence
  $(P_i)$ as shown in Fig.\ \ref{fig2}. If $d_i=OP_i$,  the  sequences
$(d_{2i})$ and $(d_{2i+1})$ are diverging. The same result holds in the $\rho_y$--$\phi_y$ subspace, and the resonator is unstable. 

\subsection{Interpretation in the scaled phase space for $-1<J=J'<0$}

We assume $L>0$, so that $\beta =\beta '$ as explained in Sect.\ \ref{sect51}. For the field transfer from ${\cal M}_1$ to ${\cal M}_2$, light is issued from ${\cal M}_1$, which is the emitter, and the radius of ${\cal M}_1$ to be taken into account is $R'_1$; and since ${\cal M}_2$ is the receiver, light is incident on ${\cal M}_2$, and the radius of ${\cal M}_2$ to be taken into account is $R_2$. The parameter ${\frak s}$ is then the sign of $R'_1(R'_1-L)$.  According to the result established in Sect.\ \ref{sect23}, the sign of $R_2(R_2+D)$ is $-{\frak s}$.

For the field transfer from ${\cal M}_2$ to ${\cal M}_1$, mirror ${\cal M}_2$ is the emitter and ${\cal M}_1$ the receiver, so that radii to be taken into account are $R'_2$ and $R_1$. The sign to be considered, denoted ${\frak s}'$, is that of $R'_2(R'_2-L)$. It is opposite to the sign of $R_1(R_1+L)$, which is $-{\frak s}'$.

Since $R_1=-R'_1$, we have $R_1(R_1+L)=R'_1(R'_1-L)$, so that ${\frak s}'=-{\frak s}$. And since
$R_2=-R'_2$, we also have $R_2(R_2+L)=R'_2(R'_2-L)$.

We assume $\beta >0$ and conclude as follows.
\begin{itemize}
\item The field transfer from ${\cal M}_1$ to ${\cal M}_2$ is described according to the sign ${\frak s}$. Its effect on the respective  Wigner distributions on ${\cal M}_1$ and ${\cal M}_2$ is a rotation of angle $-\pi /2$ followed by a hyperbolic rotation of angle $-{\frak s}\,\beta$, according to Eq.\ (\ref{eq37c}), Sect.\ \ref{sect424}.
\item The field tranfer from ${\cal M}_2$ to ${\cal M}_1$ is described according to the sign ${\frak s}'$. Its effect  on the respective  Wigner distributions on ${\cal M}_2$ and ${\cal M}_1$ is a rotation of angle $-\pi /2$ followed by a hyperbolic rotation of angle $-{\frak s}'\,\beta$. Since ${\frak s}'=-{\frak s}$, the effect of diffraction is a rotation of angle $-\pi /2$ followed by a hyperbolic rotation of angle ${\frak s}\,\beta$.
  \end{itemize}

\begin{figure}
\begin{center}
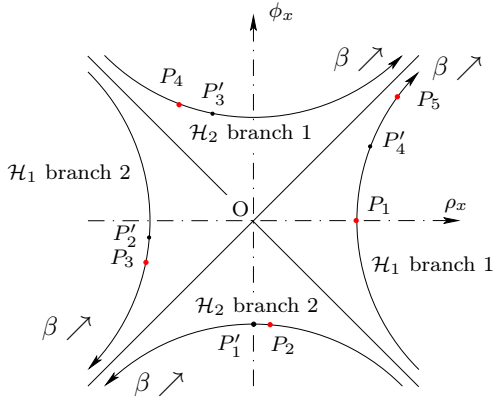
\caption{\small Case $-1<J<0$. Fractional parameter $\alpha =\pi
  /2+\I\beta$ ($\beta >0$). The sequence of points $P_i$ is analyzed in the text. If $d_i=OP_i$, sequences $(d_{2i}$) and
  $(d_{2i +1}$) are
  diverging. The corresponding  resonator is unstable.\label{fig3}
}
\end{center}
\end{figure}

The consequence for an optical resonator is illustrated by
Fig. \ref{fig3}, where $\alpha =\pi /2+\I\beta$ ($\beta >0)$ and ${\frak s}=1$ and is explained as follows.
The interpretation of point $P_i$ is that of Sect.\ \ref{sect52} once more.

Let $P_1=(\rho_{1x},\phi_{1x})$ be the initial point where the Wigner distribution on mirror ${\cal M}_1$ is considered, and let $A=\rho_{1x}^2-\phi_{1x}^2$. Let ${\cal H}_1$ be the
hyperbola whose equation is  $\rho_{x}^2-\phi_{x}^2=A$ and to which
$P_1$ belongs; and ${\cal
  H}_2$ the hyperbola whose equation is  $\phi_{x}^2-\rho_{x}^2=A$. We assume $\beta >0$, and  build a
sequence $(P_i)$ as follows.  Let $P'_1$ be the
image of $P_1$  in the elliptical (or pure) rotation of angle $-\pi /2$. Then
$P_2$   is the image of $P'_1$  in the hyperbolic rotation of parameter
$-\beta$. Both $P'_1$ and $P_2$ belong to ${\cal H}_2$. The following
back transfer from ${\cal M}_2$ to ${\cal M}_1$ transforms $P_2$ into
$P_3$, which is obtained after an elliptical rotation of angle $-\pi /2$
($P_2$ becomes $P'_2$)
and a hyperbolic rotation of parameter $\beta$ (in which $P'_2$ becomes $P_3$).

The sequence of points $P_i$ is the following (see Fig. \ref{fig3}):
\begin{enumerate}
  \item[i.] Point $P_1$ is assumed to be on ${\cal H}_1$, branch 1.
\item[ii.] First transfer from ${\cal M}_1$ to ${\cal M}_2$: from $P_1$
  to $P_2$.
\begin{itemize}
\item From $P_1$ to $P'_1$: elliptical rotation of $-\pi /2$; $P'_1\in
  {\cal H}_2$, branch 2.
\item From $P'_1$ to $P_2$: moving on ${\cal H}_2$ corresponding to a hyperbolic rotation of angle 
  $-\beta$; $P_2\in
  {\cal H}_2$, branch 2.
\end{itemize}
\item[iii.]  Back transfer from ${\cal M}_2$ to ${\cal M}_1$: from
  $P_2$ to $P_3$.
\begin{itemize}
\item From $P_2$ to $P'_2$: elliptical rotation of $-\pi /2$; $P'_2\in
  {\cal H}_1$, branch 2.
\item From $P'_2$ to $P_3$: moving on ${\cal H}_1$ corresponding to a hyperbolic rotation of angle
  $\beta$; and $P_3\in
  {\cal H}_1$, branch 2.
\end{itemize}
\item[iv.] Second transfer from ${\cal M}_1$ to ${\cal M}_2$: from
  $P_3$ to $P_4$.
\begin{itemize}
\item From $P_3$ to $P'_3$: elliptical rotation of $-\pi /2$; $P'_3\in
  {\cal H}_2$, branch 1.
\item From $P'_3$ to $P_4$: moving on ${\cal H}_2$ corresponding to
  a hyperbolic rotation of angle $-\beta$; $P_4\in
  {\cal H}_2$, branch 1.
\end{itemize}
 \goodbreak
\item[v.]  Second back transfer from ${\cal M}_2$ to ${\cal M}_1$:
  from $P_4$ to $P_5$.
\begin{itemize}
\item From $P_4$ to $P'_4$: elliptical rotation of $-\pi /2$; $P'_4\in
  {\cal H}_1$, branch 1.
\item From $P'_4$ to $P_5$: moving on ${\cal H}_1$ corresponding to
  a hyperbolic rotation of parameter $\beta$; $P_5\in
  {\cal H}_1$, branch 1.
\end{itemize}
\end{enumerate}
And so on. If $d_i=OP_i$, the sequences $(d_{2i})$ and $(d_{2i+1})$ are diverging and the considered
resonator is unstable.

\section{Conclusion}\label{conc}
Given two spherical caps, the field transfer by diffraction from one spherical cap to the other can be expressed by a fractional-order Fourier transformation. According to the distance between the spherical caps and to their curvature radii, the order of the transformation may be a real number (the field transfer is then said to be a real-order transfer) or a complex number (complex-order transfer). If the spherical caps are  the two mirrors of  an optical resonator, the order is a real number, if the resonator is stable, and a complex number, if the resonator is unstable. A unified theoretical approach can thus be developed and similarly applied to both kinds of resonators.

The effect of diffraction on the Wigner distribution associated with an optical field is shown to be a rotation, provided that the Wigner distribution is related to a scaled phase-space on which  an appropriate Euclidean structure is defined. The rotation can split into two elliptical rotations, if the field transfer is a real-order transfer; and mainly into two hyperbolic rotations, if the field transfer is a complex-order transfer. A graphical analysis clearly illustrates differences between the two kinds of resonators, according to the results of Part I and Part II of the article.

Finally, a point in the scaled phase-space corresponds to a light ray, characterized  by a point in the physical space and a direction of propagation, as will be shown in the third part of the article.  The evolution of a point in the scaled phase-space can then be interpreted like the transformation of a ray after refraction, reflection or propagation, and can be applied to 
light-ray tracing \cite{Fog4} as will be done in Part  III, with applications to ray tracing in optical resonators.

\section*{Appendix A. Proof of Eq.\ (\ref{eq20n}) ($D<0$ and $J<-1$)}\label{appenA2}

We recall that $\vec \sigma$ is an element of $\Gamma \times\Gamma  =(1+{\frak s}\,\I
){\mathbb R}\times (1+{\frak s}\,\I
){\mathbb R}$. According to Eq.\ (\ref{eq14}), for $\vec r =(x,y)$ we introduce---see Eq.\ (\ref{eq13})--- the real scaled variable $\vec\rho =(\rho_x,\rho_y)$ and the complex scaled variable $\vec \sigma
=(\sigma_x,\sigma_y)$ with
\begin{equation}\sigma_x ={1+{\frak s}\,\I\over \sqrt{2}}\,\rho_x={1+{\frak s}\,\I \over \sqrt{2\lambda \chi_1 R_1}}\,x\,,\hskip .5cm
\sigma_y = {1+{\frak s}\,\I\over \sqrt{2}}\,\rho_y={1+{\frak s}\,\I \over \sqrt{2\lambda \chi_1 R_1}}\,y\,.\end{equation}
We then obtain
\begin{equation}
  \vec\sigma\vec\cdot\vec\sigma ={\frak s}\,\I\,  \vec\rho\vec\cdot\vec\rho ={{\frak s}\,\I\over\lambda \chi_1R_1}\,\vec r\vec\cdot\vec r\,,\label{eq68}\end{equation}
and
\begin{equation}
\D\vec\sigma =\D\sigma_x\,\D\sigma_y ={\frak s}\,\I\,\D\vec \rho={{\frak s}\,\I\over  \lambda \chi_1R_1}\,\D x\,\D
y={{\frak s}\,\I\over \lambda \chi_1R_1}\,\D \vec r 
\,.\label{eq67}\end{equation}

\subsubsection*{Derivation of $\D\vec r/\lambda D$}
We start with
\begin{eqnarray}
{{\chi_1}^{\!2}{R_1}^{\!2}\over D^2}&=&{{R_1}^{\!2}\over (R_1-D)^2}\coth^2\beta =
{-{R_1}^{\!2}\over (R_1-D)^2}\cot^2\alpha 
= -{R_2+D\over R_2D}\,{R_1D\over R_1-D}\,{R_1R_2\over
  D(D-R_1+R_2)}\nonumber \\
&=& {-\chi_1R_1\over \chi_2R_2}\,{R_1R_2\over D(D-R_1+R_2)}\,.
\end{eqnarray}
Since
\begin{equation}
{\cos^2\alpha\over \sin^2\alpha}=\cot^2\alpha ={(R_1-D)(R_2+D)\over D(D-R_1+R_2)}\,,\end{equation}
we obtain
\begin{equation}
{1\over \sin^2\alpha}={R_1R_2\over D(D-R_1+R_2)}\,,
\end{equation}
and then
\begin{equation}
{\chi_1^2R_1^2\over D^2}=-{\chi_1R_1\over \chi_2R_2}{1\over
  \sin^2\alpha}={\chi_1R_1\over \chi_2R_2}{1\over \sinh^2\beta}\,.\end{equation}
Since $\chi_1R_1>0$, $\chi_2R_2>0$, since $\sinh\beta$ has the sign of
  $\beta$ which is also the sign of $D$, we may write
\begin{equation}
{\chi_1R_1\over D}={1\over \sinh\beta}\sqrt{\chi_1R_1\over
  \chi_2R_2}={\I\over \sin\alpha}\sqrt{\chi_1R_1\over \chi_2R_2}\,.\end{equation}
Finally, we use Eq. (\ref{eq67}) and write 
\begin{equation}
{\D\vec r\over \lambda D}=-{{\frak s}\,\I \,\chi_1R_1\over D}\,\D \vec \sigma ={{\frak s}\over
  \sin\alpha}\sqrt{\chi_1R_1\over
  \chi_2R_2}\,\D\vec \sigma\,.\end{equation}

\subsubsection*{Derivation of $\vec r\vec \cdot\vec r'/\lambda D$}
We use Eqs. (\ref{eq14}) and (\ref{eq14b}) and obtain
\begin{equation}
{\vec r\vec\cdot\vec r'\over\lambda D}={-\I\over \lambda
D}\sqrt{\lambda^2\chi_1R_1\chi_2R_2}\;\vec\sigma\vec\cdot\vec\sigma '
 ={-\I\over
    D}\sqrt{\chi_1R_1\chi_2R_2}\;\vec\sigma\vec\cdot\vec\sigma '\,,\end{equation}
and then
\begin{equation}
{\chi_1R_1\chi_2R_2\over D^2}= {R_1R_2\over
  (R_1-D)(R_2+D)}\,\coth^2\beta
= {-R_1R_2\over D(D-R_1+R_2)}
={-1\over \sin^2\alpha}={1\over \sinh^2\beta}\,.\end{equation}
Since $\chi_1R_1>0$ and $\chi_2R_2>0$, and since $\beta$ has the sign
  of $D$, we obtain
\begin{equation}
{\sqrt{\chi_1R_1\chi_2R_2}\over D}= {1\over \sinh\beta}={\I\over \sin\alpha}\,,\end{equation}
and then
\begin{equation}
{\vec r\vec\cdot \vec r'\over \lambda D}={\vec
  \sigma\vec\cdot\vec\sigma '\over \sin\alpha}\,.\end{equation}

\subsubsection*{Derivation of the quadratic phase terms}
We begin with
\begin{equation}
{1\over \lambda}\left({1\over D}-{1\over R_1}\right)\vec r\vec \cdot
\vec r = {-{\frak s}\,\I\over \lambda}\,{R_1-D\over DR_1}\,\lambda
\chi_1R_1\vec\sigma\vec\cdot\vec \sigma 
=-\I\,\vec \sigma\vec\cdot\vec\sigma \coth\beta
=\vec
  \sigma\vec\cdot\vec\sigma \,\cot\alpha\,,\end{equation}
and we remark that $\vec
  \sigma\vec\cdot\vec\sigma \,\cot\alpha$ is a real number.

Then
\begin{equation}
{1\over \lambda}\left({1\over D}+{1\over R_2}\right)\vec r'\vec \cdot
\vec r' = {-{\frak s}\,\I\over \lambda}\,{R_2+D\over DR_2}\,\lambda
\chi_2R_2\vec\sigma '\vec\cdot\vec \sigma '
=-\I\,\vec \sigma '\vec\cdot\vec\sigma ' \coth\beta 
=\vec
  \sigma '\vec\cdot\vec\sigma '\,\cot\alpha\,,\end{equation}
and $\vec
  \sigma '\vec\cdot\vec\sigma '\,\cot\alpha$ is also a real number.

  \subsubsection*{Integral}

The previous results lead us to write Eq.\ (\ref{eq1}) in the form
\begin{eqnarray}
U_2\left({{\frak s}-\I \over \sqrt{2}}\sqrt{\lambda
    \chi_2R_2}\;\vec\sigma '\right)\!\!\!&=&
\!\!\!{{\frak s}\,\I\over \sin\alpha} \sqrt{\chi_1R_1\over \chi_2R_2}
   \exp
    (-\I\pi \,\vec \sigma '\!\!\vec\cdot\!\vec\sigma '\cot\alpha ) \\
& & \hskip -1.5cm\times \int_{\Gamma\times  \Gamma}\!\! \!\!\!\exp (-\I\pi\, \vec\sigma\!\vec\cdot\!\vec\sigma\cot\alpha )\,
\exp\left({2\I\pi \over \sin\alpha}\,\vec\sigma \!\vec \cdot\!\vec\sigma
    '\right)\, U_1 \!\!\left({1-{\frak s}\,\I\over
  \sqrt{2}}\sqrt{\lambda\chi_1R_1}\;\vec\sigma\right)\,\D\vec
    \sigma\,,\nonumber \end{eqnarray}
that is
\begin{equation}
V_{{\rm c}2}(\vec \sigma ')= {{\frak s}\,\I\over\sin\alpha}\exp
(-\I\pi\vec\sigma '\!\!\vec\cdot\!\vec\sigma '\cot\alpha )
 \!\!\int_{{\it \Gamma}\times {\it\Gamma}} \!\!\!\!\!\exp (-\I\pi\vec\sigma\!\vec\cdot\!\vec
\sigma\cot\alpha )
\exp\!\left({2\I\pi\over\sin\alpha}\vec\sigma\! \vec\cdot\!\vec\sigma
  '\!\right)V_{{\rm c}1}(\vec\sigma )\,\D\vec \sigma \,,\label{eq94}\end{equation}
which is Eq.\ (\ref{eq20n}).

\section*{Appendix B. Proof of Eq.\ (\ref{eq20s}) ($D<0$ and $J<-1$)\label{appenB}}
This is the case  $\beta <0$ and ${\frak s}$ is the sign of $R_1(R_1-D)$. Complex scaled variables are
\begin{equation}\sigma_x ={1+{\frak s}\,\I\over \sqrt{2}}\,\rho_x={1+{\frak s}\,\I \over \sqrt{2\lambda \chi_1 R_1}}\,x\,,\hskip .5cm
  \sigma_y = {1+{\frak s}\,\I\over \sqrt{2}}\,\rho_y={1+{\frak s}\,\I \over \sqrt{2\lambda \chi_1 R_1}}\,y\,.\label{eq101n}\end{equation}
and
\begin{equation}\sigma '_x =-{{\frak s}+\I\over \sqrt{2}}\,\rho '_x=-{{\frak s}+\I \over \sqrt{2\lambda \chi_2 R_2}}\,x'\,,\hskip .5cm
  \sigma '_y = -{{\frak s}+\I\over \sqrt{2}}\,\rho '_y=-{{\frak s}+\I \over \sqrt{2\lambda \chi_2 R_2}}\,y'\,.\label{eq102n}\end{equation}

We then obtain
\begin{equation}
  \vec\sigma\vec\cdot\vec\sigma ={\frak s}\,\I\,  \vec\rho\vec\cdot\vec\rho ={{\frak s}\,\I\over\lambda \chi_1R_1}\,\vec r\vec\cdot\vec r\,,\label{eq68a}\end{equation}
and
\begin{equation}
\D\vec\sigma =\D\sigma_x\,\D\sigma_y ={\frak s}\,\I\,\D\vec \rho={{\frak s}\,\I\over  \lambda \chi_1R_1}\,\D x\,\D
y={{\frak s}\,\I\over \lambda \chi_1R_1}\,\D \vec r 
\,.\label{eq67a}\end{equation}

\subsubsection*{Derivation of $\D\vec r/\lambda D$}
We start with
\begin{eqnarray}
{\chi_1^2R_1^2\over D^2}&=&{R_1^2\over (R_1-D)^2}\coth^2\beta =
{-R_1^2\over (R_1-D)^2}\cot^2\alpha 
= -{R_2+D\over R_2D}\,{R_1D\over R_1-D}\,{R_1R_2\over
  D(D-R_1+R_2)}\nonumber \\
&=& {-\chi_1R_1\over \chi_2R_2}\,{R_1R_2\over D(D-R_1+R_2)}\,.
\end{eqnarray}
Since
\begin{equation}
{\cos^2\alpha\over \sin^2\alpha}=\cot^2\alpha ={(R_1-D)(R_2+D)\over D(D-R_1+R_2)}\,,\end{equation}
we obtain
\begin{equation}
{1\over \sin^2\alpha}={R_1R_2\over D(D-R_1+R_2)}\,,
\end{equation}
and then
\begin{equation}
{\chi_1^2R_1^2\over D^2}=-{\chi_1R_1\over \chi_2R_2}{1\over
  \sin^2\alpha}={\chi_1R_1\over \chi_2R_2}{1\over \sinh^2\beta}\,.\end{equation}
Since $\chi_1R_1>0$, $\chi_2R_2>0$, since $\sinh\beta$ has the sign of
  $\beta$ which is also the sign of $D$, we may write
\begin{equation}
{\chi_1R_1\over D}={1\over \sinh\beta}\sqrt{\chi_1R_1\over
  \chi_2R_2}={\I\over \sin\alpha}\sqrt{\chi_1R_1\over \chi_2R_2}\,.\end{equation}
Finally, we use Eq. (\ref{eq67}) and write 
\begin{equation}
{\D\vec r\over \lambda D}={-{\frak s}\,\I \,\chi_1R_1\over D}\,\D \vec \sigma ={{\frak s}\over
  \sin\alpha}\sqrt{\chi_1R_1\over
  \chi_2R_2}\,\D\vec \sigma\,.\end{equation}

\subsubsection*{Derivation of $\vec r\vec \cdot\vec r'/\lambda D$}
We use Eqs. (\ref{eq101n}) and (\ref{eq102n}) and obtain
\begin{equation}
{\vec r\vec\cdot\vec r'\over\lambda D}={\I\over \lambda
D}\sqrt{\lambda^2\chi_1R_1\chi_2R_2}\;\vec\sigma\vec\cdot\vec\sigma '
 ={\I\over
    D}\sqrt{\chi_1R_1\chi_2R_2}\;\vec\sigma\vec\cdot\vec\sigma '\,,\end{equation}
and then
\begin{equation}
{\chi_1R_1\chi_2R_2\over D^2}= {R_1R_2\over
  (R_1-D)(R_2+D)}\,\coth^2\beta
= {-R_1R_2\over D(D-R_1+R_2)}
={-1\over \sin^2\alpha}={1\over \sinh^2\beta}\,.\end{equation}
Since $\chi_1R_1>0$ and $\chi_2R_2>0$, and since $\beta$ has the sign
  of $D$, we obtain
\begin{equation}
{\sqrt{\chi_1R_1\chi_2R_2}\over D}= {1\over \sinh\beta}={\I\over \sin\alpha}\,,\end{equation}
and then
\begin{equation}
{\vec r\vec\cdot \vec r'\over \lambda D}=-{\vec
  \sigma\vec\cdot\vec\sigma '\over \sin\alpha}\,.\end{equation}

\subsubsection*{Derivation of the quadratic phase terms}
We begin with
\begin{equation}
{1\over \lambda}\left({1\over D}-{1\over R_1}\right)\vec r\vec \cdot
\vec r = {-{\frak s}\,\I\over \lambda}\,{R_1-D\over DR_1}\,\lambda
\chi_1R_1\vec\sigma\vec\cdot\vec \sigma 
=-\I\,\vec \sigma\vec\cdot\vec\sigma \coth\beta
=\vec
  \sigma\vec\cdot\vec\sigma \,\cot\alpha\,,\end{equation}
and we remark that $\vec
  \sigma\vec\cdot\vec\sigma \,\cot\alpha$ is a real number.

Then
\begin{equation}
{1\over \lambda}\left({1\over D}+{1\over R_2}\right)\vec r'\vec \cdot
\vec r' = {-{\frak s}\,\I\over \lambda}\,{R_2+D\over DR_2}\,\lambda
\chi_2R_2\vec\sigma '\vec\cdot\vec \sigma '
=-\I\,\vec \sigma '\vec\cdot\vec\sigma ' \coth\beta 
=\vec
  \sigma '\vec\cdot\vec\sigma '\,\cot\alpha\,,\end{equation}
and $\vec
  \sigma '\vec\cdot\vec\sigma '\,\cot\alpha$ is also a real number.

  \subsubsection*{Integral}

The previous results lead us to write Eq. (\ref{eq1}) in the form
\begin{eqnarray}
U_2\left({-{\frak s}+\I \over \sqrt{2}}\sqrt{\lambda
    \chi_2R_2}\;\vec\sigma '\right)\!\!\!&=&
\!\!\!{-{\frak s}\,\I\over \sin\alpha} \sqrt{\chi_1R_1\over \chi_2R_2}
   \exp
    (\I\pi \,\vec \sigma '\!\!\vec\cdot\!\vec\sigma '\cot\alpha ) \\
& & \hskip -1.5cm\times \int_{\Gamma\times  \Gamma} \!\!\!\exp (\I\pi\, \vec\sigma\!\vec\cdot\!\vec\sigma\cot\alpha )\,
\exp\left({-2\I\pi \over \sin\alpha}\,\vec\sigma \!\vec \cdot\!\vec\sigma
    '\right)\, U_1 \!\!\left({1-{\frak s}\,\I\over
  \sqrt{2}}\sqrt{\lambda\chi_1R_1}\;\vec\sigma\right)\,\D\vec
    \sigma\,,\nonumber \end{eqnarray}
that is
\begin{equation}
V_{{\rm c}2}(\vec \sigma ')= {-{\frak s}\,\I\over\sin\alpha}\exp
(\I\pi\vec\sigma '\!\!\vec\cdot\!\vec\sigma '\cot\alpha )
 \!\!\int_{{\it \Gamma}\times {\it\Gamma}} \!\!\!\!\!\exp (\I\pi\vec\sigma\!\vec\cdot\!\vec
\sigma\cot\alpha )
\exp\!\left(-{2\I\pi\over\sin\alpha}\vec\sigma\! \vec\cdot\!\vec\sigma
  '\!\right)V_{{\rm c}1}(\vec\sigma )\,\D\vec \sigma \,,\label{eq94a}\end{equation}
which is Eq.\ (\ref{eq20s}).

\section*{Appendix C. Proof of Eq.\ (\ref{eq20nb}) ($D>0$ and $-1<J<0$)\label{appenC}}
We have $\alpha =(\pi /2)+\I\beta$, where $\beta>0$, since it  has the sign of $D$. Then $\cos\alpha =-\I\sin\I\beta=\sinh\beta>0$, and $\sin\alpha =\cos\I\beta=\cosh\beta >0$.

Complex scaled-variables are
\begin{equation}\sigma_x ={1+\I\over \sqrt{2}}\,\rho_x={1+\I \over \sqrt{2\lambda \chi_1 R_1}}\,x\,,\hskip .5cm
  \sigma_y = {1+\I\over \sqrt{2}}\,\rho_y={1+\I \over \sqrt{2\lambda \chi_1 R_1}}\,y\,,\label{eq101nc}\end{equation}
and
\begin{equation}\sigma '_x ={1-\I\over \sqrt{2}}\,\rho '_x={1-\I \over \sqrt{2\lambda \chi_2 R_2}}\,x'\,,\hskip .5cm
  \sigma '_y = {1-\I\over \sqrt{2}}\,\rho '_y={1-\I \over \sqrt{2\lambda \chi_2 R_2}}\,y'\,.\label{eq102nc}\end{equation}

We then obtain
\begin{equation}
  \vec\sigma\vec\cdot\vec\sigma =\I\,  \vec\rho\vec\cdot\vec\rho ={\I\over\lambda \chi_1R_1}\,\vec r\vec\cdot\vec r\,,\label{eq68c}\end{equation}
and
\begin{equation}
\D\vec\sigma =\D\sigma_x\,\D\sigma_y =\I\,\D\vec \rho={\I\over  \lambda \chi_1R_1}\,\D x\,\D
y={\I\over \lambda \chi_1R_1}\,\D \vec r
\,.\label{eq67c}\end{equation}

\subsubsection*{Derivation of $\D\vec r/\lambda D$}
From Eq.\ (\ref{eq21}) we deduce 
\begin{eqnarray}
  {\chi_1^2R_1^2\over D^2}&=&{R_1^2\over (R_1-D)^2}\,{1\over \coth^2\beta} =
  -J{R_1^2\over (R_1-D)^2}
=  - {{R_1}^2\over D(D-R_1+R_2)}\,{R_2+D\over R_1-D}
\nonumber \\
&=& {\chi_1R_1\over \chi_2R_2}\,{R_1R_2\over D(D-R_1+R_2)}\,.
\end{eqnarray}
Since
\begin{equation}
{\cos^2\alpha\over \sin^2\alpha}=\cot^2\alpha ={(R_1-D)(R_2+D)\over D(D-R_1+R_2)}\,,\end{equation}
we obtain
\begin{equation}
{1\over \sin^2\alpha}={R_1R_2\over D(D-R_1+R_2)}\,,
\end{equation}
and then
\begin{equation}
{\chi_1^2R_1^2\over D^2}={\chi_1R_1\over \chi_2R_2}{1\over
  \sin^2\alpha}={\chi_1R_1\over \chi_2R_2}{1\over \cosh^2\beta}\,.\end{equation}
Since $\chi_1R_1>0$, $\chi_2R_2>0$, since $\sinh\beta$ has the sign of
  $\beta$ which is also the sign of $D$, we may write
\begin{equation}
{\chi_1R_1\over D}={1\over \cosh\beta}\sqrt{\chi_1R_1\over
  \chi_2R_2}={1\over \sin\alpha}\sqrt{\chi_1R_1\over \chi_2R_2}\,.\end{equation}
Finally, we use Eq. (\ref{eq67c}) and write 
\begin{equation}
{\D\vec r\over \lambda D}={-\I \,\chi_1R_1\over D}\,\D \vec \sigma ={-\I\over
  \sin\alpha}\sqrt{\chi_1R_1\over
  \chi_2R_2}\,\D\vec \sigma\,.\end{equation}

\subsubsection*{Derivation of $\vec r\vec \cdot\vec r'/\lambda D$}
We start with 
\begin{equation}
{\vec r\vec\cdot\vec r'\over\lambda D}={1\over \lambda
D}\sqrt{\lambda^2\chi_1R_1\chi_2R_2}\;\vec\sigma\vec\cdot\vec\sigma '
 ={1\over
    D}\sqrt{\chi_1R_1\chi_2R_2}\;\vec\sigma\vec\cdot\vec\sigma '\,,\end{equation}
and we use Eqs. (\ref{eq21}) and (\ref{eq23}) to obtain
\begin{equation}
{\chi_1R_1\chi_2R_2\over D^2}= {-R_1R_2\over
  (R_1-D)(R_2+D)}\,{1\over \coth^2\beta}
= {R_1R_2\over D(D-R_1+R_2)}
={1\over \sin^2\alpha}={1\over \cosh^2\beta}\,.\end{equation}
Since $\chi_1R_1>0$ and $\chi_2R_2>0$, and since $D>0$, we obtain
\begin{equation}
{\sqrt{\chi_1R_1\chi_2R_2}\over D}= {1\over \cosh\beta}={1\over \sin\alpha}\,,\end{equation}
and then
\begin{equation}
{\vec r\vec\cdot \vec r'\over \lambda D}={\vec
  \sigma\vec\cdot\vec\sigma '\over \sin\alpha}\,.\end{equation}

\subsubsection*{Derivation of the quadratic phase terms}
We begin with
\begin{equation}
{1\over \lambda}\left({1\over D}-{1\over R_1}\right)\vec r\vec \cdot
\vec r = {-\I\over \lambda}\,{R_1-D\over DR_1}\,\lambda
\chi_1R_1\,\vec\sigma\vec\cdot\vec \sigma 
={-{\frak s}\,\I\,\vec \sigma\vec\cdot\vec\sigma\over  \coth\beta}
={\frak s}\,\vec
  \sigma\vec\cdot\vec\sigma \,\cot\alpha\,,\end{equation}
and we remark that $\vec
  \sigma\vec\cdot\vec\sigma \,\cot\alpha$ is a real number.

Then
\begin{equation}
{1\over \lambda}\left({1\over D}+{1\over R_2}\right)\vec r'\vec \cdot
\vec r' = {\I\over \lambda}\,{R_2+D\over DR_2}\,\lambda
\chi_2R_2\,\vec\sigma '\vec\cdot\vec \sigma '
={-{\frak s}\,\I\,\vec \sigma '\vec\cdot\vec\sigma '\over  \coth\beta} 
={\frak s}\,\vec
  \sigma '\vec\cdot\vec\sigma '\,\cot\alpha\,,\end{equation}
and $\vec
  \sigma '\vec\cdot\vec\sigma '\,\cot\alpha$ is also a real number.

  \subsubsection*{Integral}

The previous results lead us to write Eq. (\ref{eq1}) in the form
\begin{eqnarray}
U_2\left({1+\I \over \sqrt{2}}\sqrt{\lambda
    \chi_2R_2}\;\vec\sigma '\right)\!\!\!&=&
\!\!\!{-\I\over \sin\alpha} \sqrt{\chi_1R_1\over \chi_2R_2}
   \exp
    (-{\frak s}\,\I\pi \,\vec \sigma '\!\!\vec\cdot\!\vec\sigma '\cot\alpha ) \\
& & \hskip -1.5cm\times \int_{\Gamma\times  \Gamma} \!\!\!\exp (-{\frak s}\,\I\pi\, \vec\sigma\!\vec\cdot\!\vec\sigma\cot\alpha )\,
\exp\left({2\I\pi \over \sin\alpha}\,\vec\sigma \!\vec \cdot\!\vec\sigma
    '\right)\, U_1 \!\!\left({1-\I\over
  \sqrt{2}}\sqrt{\lambda\chi_1R_1}\;\vec\sigma\right)\,\D\vec
    \sigma\,,\nonumber \end{eqnarray}
that is
\begin{eqnarray}
V_{{\rm c}2}(\vec \sigma ')&=& {-\I\over\sin\alpha}\exp
(-{\frak s}\,\I\pi\vec\sigma '\!\!\vec\cdot\!\vec\sigma '\cot\alpha )\nonumber \\
& &\hskip 1cm \times \int_{{\Gamma}\times {\Gamma}} \!\!\!\!\!\exp (-{\frak s}\,\I\pi\vec\sigma\!\vec\cdot\!\vec
\sigma\cot\alpha )
\exp\!\left({2\I\pi\over\sin\alpha}\vec\sigma\! \vec\cdot\!\vec\sigma
  '\!\right)V_{{\rm c}1}(\vec\sigma )\,\D\vec \sigma \,,\label{eq94c}\end{eqnarray}
which  is Eq.\ (\ref{eq20nb}).



\begin{thebibliography}{references}\label{refe2}

\leftskip = -1cm


  \bibitem{Part1}  P. Pellat-Finet, \'E. Fogret, ``Effect of Diffraction on Wigner Distributions  of 
    Optical  Fields and how to Use It in Optical Resonator Theory. I -- Stable Resonators and Gaussian Beams,'' arXiv 2005.13430v1 (2020) 1--20. (2005.13430v2: 2022.)

    \bibitem{PPF5} P. Pellat-Finet, {\sl Optique de Fourier. Théorie métaxiale
  et fractionnaire}, Springer, Paris, 2009.


\bibitem{Fog1}  P. Pellat-Finet, \'E. Fogret, ``Complex order fractional
  Fourier transforms and their use in diffraction theory,'' Opt. Comm. {\bf 258}
(2006) 103--113.



\bibitem{Fog3} {P. Pellat-Finet, \'E. Fogret}, ``Fractional Fourier optics theory of
  optical resonators,'' in: P.~S. Emersone (Ed.), {\sl Progress in optical fibers}, Nova  Science Publishers, New York (2011)  299--351.

  

  \bibitem{Nam} {V. Namias}, ``The fractional order Fourier transform and
its applications to quantum mechanics,'' J. Inst. Maths Applics {\bf 25}
(1980) 241--265.

\bibitem{Mcb} {A. C. McBride, F. H. Kerr}, ``On Namias's fractional Fourier transform,'' IMA J. Appl. Math. {\bf 39} (1987) 159--175.

  \bibitem{PPF4b}  P. Pellat-Finet, P.-E. Durand, \'E. Fogret, ``Spherical
 angular spectrum and the fractional order Fourier transform,'' Opt. Lett.  {\bf
    31} (2006) 3429--3431.

  \bibitem{Ana} {Y. A. Anan'ev},  {\sl Optical resonators and the beam divergence problem}, IOP, Bristol, 1992.

\bibitem{Sie2} {A. E. Siegman.}, {\sl Lasers}, University Science Books, Mill Valley, 1986.


\bibitem{Fog4} {P. Pellat-Finet, \'E. Fogret}, ``Ray tracing based on the Wigner representation of optical fields,'' 
\'Optica Pura y Aplicada  {\bf 51} (2018) 49025:1-10.

\end{thebibliography}
\end{document}